\documentclass[twocolumn]{aastex631}

\usepackage{savesym}
\savesymbol{tablenum}
\usepackage{siunitx}
\restoresymbol{SIX}{tablenum}

\usepackage[version=4]{mhchem}
\usepackage{graphicx,epsf}
\usepackage{subfigure}
\usepackage{graphicx}	
\usepackage{amsmath}	
\usepackage{amssymb}	
\usepackage{newtxtext,newtxmath}
\usepackage{siunitx}
\usepackage{footnote}
\usepackage{longtable,gensymb}
\usepackage[T1]{fontenc}

\usepackage[maxfloats=256]{morefloats}

\def\arcsec{\hbox{$^{\prime\prime}$}}

\def\AT{AT\,2024wpp}

\DeclareRobustCommand{\ion}[2]{\relax\ifmmode\ifx\testbx\f@series{\mathbf{#1\,\mathsc{#2}}}\else{\mathrm{#1\,\mathsc{#2}}}\fi\else\textup{#1\,{\mdseries\textsc{#2}}}\fi}
\shorttitle{AT\,2024wpp in the UVOIR}
\shortauthors{LeBaron et al.}

\begin{document}
\correspondingauthor{Natalie LeBaron}
\email{nlebaron@berkeley.edu}

\author[0000-0002-2249-0595]{Natalie LeBaron}
\affiliation{Department of Astronomy, University of California, Berkeley, CA 94720-3411, USA}
\affiliation{Berkeley Center for Multi-messenger Research on Astrophysical Transients and Outreach (Multi-RAPTOR), University of California, Berkeley, CA 94720-3411, USA}

\author[0000-0003-4768-7586]{Raffaella Margutti}
\affiliation{Department of Astronomy, University of California, Berkeley, CA 94720-3411, USA}
\affiliation{Department of Physics, University of California, 366 Physics North MC 7300, Berkeley, CA 94720, USA}
\affiliation{Berkeley Center for Multi-messenger Research on Astrophysical Transients and Outreach (Multi-RAPTOR), University of California, Berkeley, CA 94720-3411, USA}

\author[0000-0002-7706-5668]{Ryan Chornock}
\affiliation{Department of Astronomy, University of California, Berkeley, CA 94720-3411, USA}
\affiliation{Berkeley Center for Multi-messenger Research on Astrophysical Transients and Outreach (Multi-RAPTOR), University of California, Berkeley, CA 94720-3411, USA}

\author[0000-0002-8070-5400]{Nayana A.~J.} 
\affiliation{Department of Astronomy, University of California, Berkeley, CA 94720-3411, USA}
\affiliation{Berkeley Center for Multi-messenger Research on Astrophysical Transients and Outreach (Multi-RAPTOR), University of California, Berkeley, CA 94720-3411, USA}

\author[0000-0001-5674-8403]{Olivia~Aspegren}
\affiliation{Department of Astronomy, University of California, Berkeley, CA 94720-3411, USA}

\author[0000-0002-1568-7461]{Wenbin~Lu} 
\affiliation{Department of Astronomy, University of California, Berkeley, CA 94720-3411, USA}
\affiliation{Theoretical Astrophysics Center, UC Berkeley, Berkeley, CA 94720, USA}
\affiliation{Berkeley Center for Multi-messenger Research on Astrophysical Transients and Outreach (Multi-RAPTOR), University of California, Berkeley, CA 94720-3411, USA}

\author[0000-0002-4670-7509]{Brian~D.~Metzger}
\affiliation{Department of Physics and Columbia Astrophysics Laboratory, Columbia University, New York, NY 10027, USA}
\affiliation{Center for Computational Astrophysics, Flatiron Institute, 162 5th Ave, New York, NY 10010, USA}

\author[0000-0002-5981-1022]{Daniel Kasen}
\affiliation{Department of Physics and Astronomy, University of California, Berkeley, CA 94720, USA} 
\affiliation{Nuclear Science Division, Lawrence Berkeley National Laboratory, 1 Cyclotron Road, Berkeley, CA 94720, USA}
\affiliation{Berkeley Center for Multi-messenger Research on Astrophysical Transients and Outreach (Multi-RAPTOR), University of California, Berkeley, CA 94720-3411, USA}

\author[0000-0001-5955-2502]{Thomas~G.~Brink}
\affiliation{Department of Astronomy, University of California, Berkeley, CA 94720-3411, USA}

\author[0000-0001-6278-1576]{Sergio~Campana} 
\affiliation{INAF—Osservatorio Astronomico di Brera, Via Bianchi 46, I-23807 Merate (LC), Italy}

\author[0000-0001-7164-1508]{Paolo D'Avanzo} 
\affiliation{INAF—Osservatorio Astronomico di Brera, Via Bianchi 46, I-23807 Merate (LC), Italy}

\author[0000-0001-9855-5781]{Jakob T. Faber}
\affiliation{Cahill Center for Astronomy and Astrophysics, MC 249-17 California Institute of Technology, Pasadena CA 91125, USA}

\author[0009-0007-5708-7978]{Matteo Ferro} 
\affiliation{Università degli Studi dell’Insubria, Dipartimento di Scienza e Alta Tecnologia, Via Valleggio 11, 22100 Como, Italy}
\affiliation{INAF—Osservatorio Astronomico di Brera, Via Bianchi 46, I-23807 Merate (LC), Italy}

\author[0000-0003-3460-0103]{Alexei~V.~Filippenko}
\affiliation{Department of Astronomy, University of California, Berkeley, CA 94720-3411, USA}

\author[0000-0002-2445-5275]{Ryan~J.~Foley}
\affiliation{Department of Astronomy and Astrophysics, University of California, Santa Cruz, CA 95064, USA}

\author[0009-0002-9727-8326]{Xinze Guo}
\affiliation{Department of Astronomy, University of California, Berkeley, CA 94720-3411, USA}

\author[0000-0002-5698-8703]{Erica Hammerstein} \affiliation{Department of Astronomy, University of California, Berkeley, CA 94720-3411, USA}

\author[0000-0001-8738-6011]{Saurabh~W.~Jha}
\affiliation{Department of Physics and Astronomy, Rutgers, the State University of New Jersey, 136 Frelinghuysen Road, Piscataway, NJ 08854-8019, USA}

\author[0000-0002-5740-7747]{Charles D. Kilpatrick}
\affiliation{Center for Interdisciplinary Exploration and Research in Astrophysics (CIERA) and Department of Physics and Astronomy, Northwestern University, Evanston, IL 60208, USA}

\author[0000-0002-0786-7307]{Giulia~Migliori}
\affiliation{INAF Istituto di Radioastronomia, via Gobetti 101, 40129 Bologna, Italy}

 \author[0000-0002-0763-3885]{Dan Milisavljevic}
\affiliation{Department of Physics and Astronomy, Purdue University, 525 Northwestern Ave, West Lafayette, IN 47907, USA}
\affiliation{Institute of Physical Artificial Intelligence, Purdue University, West Lafayette, IN 47907, USA}

\author[0000-0002-1092-6806]{Kishore C. Patra}
\affiliation{Department of Astronomy and Astrophysics, University of California, Santa Cruz, CA 95064, USA}

\author[0000-0001-8023-4912]{Huei Sears}
\affiliation{Department of Physics and Astronomy, Rutgers, the State University of New Jersey, 136 Frelinghuysen Road, Piscataway, NJ 08854-8019, USA}

\author[0000-0002-9486-818X]{Jonathan J. Swift}
\affiliation{The Thacher School, 5025 Thacher Rd., Ojai, CA 93023, USA}

\author[0000-0002-1481-4676]{Samaporn~Tinyanont}
\affiliation{National Astronomical Research Institute of Thailand, 260 Moo 4, Donkaew, Maerim, Chiang Mai, 50180, Thailand}

\author[0000-0002-7252-5485]{Vikram Ravi} 
\affiliation{Cahill Center for Astronomy and Astrophysics, MC 249-17 California Institute of Technology, Pasadena CA 91125, USA}
\affiliation{Owens Valley Radio Observatory, California Institute of Technology, Big Pine, CA 93513, USA}

\author[0000-0001-6747-8509]{Yuhan~Yao}
\affiliation{Miller Institute for Basic Research in Science, 206B Stanley Hall, Berkeley, CA 94720, USA} 
\affiliation{Department of Astronomy, University of California, Berkeley, CA 94720-3411, USA}
\affiliation{Berkeley Center for Multi-messenger Research on Astrophysical Transients and Outreach (Multi-RAPTOR), University of California, Berkeley, CA 94720-3411, USA}

\author[0000-0002-8297-2473]{Kate~D.~Alexander}
\affiliation{Department of Astronomy/Steward Observatory, 933 North Cherry Avenue, Rm. N204, Tucson, AZ 85721-0065, USA}

\author[0000-0002-6688-3307]{Prasiddha~Arunachalam} 
\affiliation{Department of Astronomy \& Astrophysics, University of California, Santa Cruz, CA 95064, USA}

\author[0000-0002-9392-9681]{Edo~Berger}
\affiliation{Center for Astrophysics \textbar{} Harvard \& Smithsonian, 60 Garden Street, Cambridge, MA 02138-1516, USA}

\author[0000-0002-7735-5796]{Joe S. Bright}
\affiliation{Astrophysics, Department of Physics, University of Oxford, Keble Road, Oxford OX1 3RH, UK}
\affiliation{Breakthrough Listen, Astrophysics, Department of Physics, University of Oxford, Keble Road, Oxford OX1 3RH, UK}

\author{Chuck Cynamon}
\affiliation{Supra Solem Observatory, Santa Lucia Mountains, CA, USA}

\author[0000-0002-5680-4660]{Kyle~W.~Davis}
\affiliation{Department of Astronomy and Astrophysics, University of California, Santa Cruz, CA 95064, USA}

\author[0000-0001-6922-8319]{Braden Garretson}
\affiliation{Department of Physics and Astronomy, Purdue University, 525 Northwestern Avenue, West Lafayette, IN 47907, USA}

\author[0000-0001-8867-4234]{Puragra~Guhathakurta} 
\affiliation{Department of Astronomy and Astrophysics, University of California, Santa Cruz, CA 95064, USA}

\author[0000-0002-3934-2644]{Wynn~V.~Jacobson-Gal\'{a}n}
\altaffiliation{NASA Hubble Fellow}
\affiliation{Cahill Center for Astrophysics, California Institute of Technology, MC 249-17, 1216 E California Boulevard, Pasadena, CA, 91125, USA}

\author[0000-0002-6230-0151]{D.~O.~Jones}
\affiliation{Gemini Observatory, NSF’s NOIRLab, 670 N. A’ohoku Place, Hilo, HI 96720, USA}

\author[0009-0005-1871-7856]{Ravjit~Kaur} 
\affiliation{Department of Astronomy \& Astrophysics, University of California, Santa Cruz, CA 95064, USA}

\author{Stefan~Kimura} 
\affiliation{Physics Department, Willamette University, Salem, OR, USA}

\author[0000-0003-1792-2338]{Tanmoy Laskar}
\affiliation{Department of Physics \& Astronomy, University of Utah, Salt Lake City, UT 84112, USA}
\affiliation{Department of Astrophysics/IMAPP, Radboud University, P.O. Box 9010, 6500 GL, Nijmegen, The Netherlands}

\author{Morgan~Nuñez} 
\affiliation{Department of Physics \& Astronomy, San Francisco State University, San Francisco, CA 94132 USA}

\author[0009-0002-5096-1689]{Michaela Schwab}
\affiliation{Department of Physics and Astronomy, Rutgers, the State University of New Jersey, 136 Frelinghuysen Road, Piscataway, NJ 08854-8019, USA}

\author[0000-0001-6360-992X]{Monika~D.~Soraisam}
\affiliation{Gemini Observatory/NSF NOIRLab, 670 N.\, A'ohoku Place, Hilo, HI 96720, USA}

\author[0000-0001-7266-930X]{Nao~Suzuki} 
\affiliation{E.O. Lawrence Berkeley National Laboratory, 1 Cyclotron Road, Berkeley, CA 94720, USA}
\affiliation{Department of Physics, Florida State University, 77 Chieftan Way, Tallahassee, FL 32306, USA}

\author[0000-0002-5748-4558]{Kirsty~Taggart}
\affiliation{Department of Astronomy and Astrophysics, University of California, Santa Cruz, CA 95064, USA}

\author[0009-0002-4843-2913]{Eli~Wiston} 
\affiliation{Department of Astronomy, University of California, Berkeley, CA 94720-3411, USA}
\affiliation{Berkeley Center for Multi-messenger Research on Astrophysical Transients and Outreach (Multi-RAPTOR), University of California, Berkeley, CA 94720-3411, USA}

\author[0000-0002-6535-8500]{Yi Yang}
\affiliation{Department of Physics, Tsinghua University, Qinghua Yuan, Beijing 100084, China}

\author[0000-0002-2636-6508]{WeiKang Zheng}
\affiliation{Department of Astronomy, University of California, Berkeley, CA 94720-3411, USA}

\title{The Most Luminous Known Fast Blue Optical Transient AT\,2024wpp:  Unprecedented Evolution and Properties in the Ultraviolet to the Near-Infrared}

\begin{abstract}
We present an extensive photometric and spectroscopic ultraviolet-optical-infrared campaign on the luminous fast blue optical transient (LFBOT) \AT{}\, over the first $\sim100$\,d. \AT{} is the most luminous LFBOT discovered to date, with $L_{\rm{pk}}\approx(2-4)\times10^{45}\,\rm{erg\,s^{-1}}$ (5--10 times that of the prototypical AT\,2018cow). This extreme luminosity enabled the acquisition of the most detailed LFBOT UV light curve thus far. In the first $\sim45\,$d, \AT{}\, radiated $>10^{51}\,\rm{erg}$, surpassing AT\,2018cow by an order of magnitude and requiring a power source beyond the radioactive $^{56}$Ni decay of traditional supernovae. Like AT\,2018cow, the UV-optical spectrum of \AT\, is dominated by a persistently blue thermal continuum throughout our monitoring, with blackbody parameters at peak of $T>30{,}000\,$K and $R_{\rm{BB}}/t\approx0.2-0.3c$. A temperature of $\gtrsim20{,}000\,$K is maintained thereafter without evidence for cooling. We interpret the featureless spectra as a consequence of continuous energy injection from a central source of high-energy emission which maintains high ejecta ionization. After $35\,$d, faint (equivalent width $\lesssim10$\,\AA) H and He spectral features with kinematically separate velocity components centered at $0\,\rm{km\,s}^{-1}$ and $-6400\,\rm{km\,s}^{-1}$ emerge, implying spherical symmetry deviations. A near-infrared excess of emission above the optical blackbody emerges between 20--30\,d with a power-law spectrum $F_{\rm\nu,NIR}\propto\nu^{-0.3}$ at 30\,d. We interpret this distinct emission component as either reprocessing of early UV emission in a dust echo or free-free emission in an extended medium above the optical photosphere. LFBOT asphericity and multiple outflow components (including mildly relativistic ejecta) together with the large radiated energy are naturally realized by super-Eddington accretion disks around neutron stars or black holes and their outflows.

\end{abstract}



\section{Introduction} \label{sec:intro} 
Recently, high-cadence, wide-field optical transient surveys have led to the identification of a new class of astrophysical phenomena, Fast Blue Optical Transients (FBOTs). Characterized by an extremely rapid rise to maximum light ($t_{\rm rise}\lesssim 10$\,d), luminous emission which can reach $L_{\rm pk}>10^{45}\,\rm{erg\,s^{-1}}$, and persistent blue colors for weeks after peak \citep{drout2014aj,Tanaka2016ApJ, Pursiainen2018MNRAS, Arcavi2016ApJ, nicholl_at_2023,rest2018na,prentice_cow_2018, ho2023a}, these transients challenge traditional supernova (SN) models that are powered by the radioactive decay of $^{56}$Ni. Alternative sources of energy are therefore needed. Proposed sources include shock interaction with dense circumstellar material (CSM; e.g., \citealt{pellegrino2022aj, Margalit2022ApJ, Khatami2024ApJ...972..140K}) and a central engine powered either by magnetar spin-down or accretion onto a compact object.  There is debate on the nature of the compact object as well as (in the latter case) on the origin of the accreted material. Considered models include magnetar powered supernovae \citep[SNe;][]{prentice_cow_2018, vurm2021aj}, accretion onto a compact object at the center of a failed SN \citep{Margutti19,quataert2019mnras}, merger of a black hole (BH) and Wolf–Rayet (WR) star \citep{metzger2022aj}, tidal disruption event (TDE) of a main-sequence companion star by a stellar-mass BH or neutron star (NS; \citealt{tsuna2025}), and TDE by an intermediate-mass BH \citep[IMBH;][]{perley2019mnras, Gutierrez2024, ho2023_2022tsd}.

FBOTs span a wide range of peak luminosities ($L_{\rm{pk}}\approx10^{42}-10^{45}\,\rm{erg\,s}^{-1}$; e.g., \citealt{ho2023a}), are not intrinsically rare (7--11\% of the core-collapse (CC)SN rate; \citealt{drout2014aj,ho2023a}), and are likely a heterogeneous class. Lower luminosity ($L_{\rm{pk}}\lesssim 10^{43}\,\rm{erg\,s}^{-1}$) FBOTs  likely represent manifestations of the fast-evolving tail of hydrogen-poor SNe (e.g.,~\citealt{ho2023a}) possibly powered by shock interaction (and subsequent shock-cooling emission) between their fast ejecta and surrounding CSM. However, there is a subset of FBOTs with $L_{\rm{pk}}>10^{43}\,\rm{erg\,s}^{-1}$ that are also associated with luminous
radio and/or X-ray emission. These luminous (L)FBOTs (also called ``cow-like'' transients after the prototypical AT\,2018cow) are much rarer ($<$\,1\% the CCSN rate; \citealt{coppejans2020aj, ho2023a}), and thus either represent the outcome of a more unusual stellar evolution pathway (e.g., \citealt{tsuna2025}) or might not be stellar explosions at all (e.g., \citealt{metzger2022aj}). Their persistent blue colors (indicative of high temperatures $>10^{4}\,$K) and featureless ultraviolet (UV)--optical--near-infrared (NIR) spectra over many weeks point to the presence of a central heating source that is distinct from the outer shock CSM interaction. From this perspective LFBOTs present clear observational analogies to the recently identified class of featureless TDEs \citep{Hammerstein2023a, Yao2023ApJ...955L...6Y}, which may also extend to the underlying physics of these phenomena. 

Identified LFBOTs accompanied with X-ray/radio emission include AT\,2018cow \citep{Margutti19,perley2019mnras, prentice_cow_2018, chen2023a, Ho2019ApJ...871...73H}, AT\,2018lug (ZTF18abvkwla, the ``Koala''; \citealt{ho2020aja}), AT\,2020mrf \citep{yao2022aj}, AT\,2022tsd \citep{matthews2023rnaas,ho2023_2022tsd}, CRTS-CSS161010 J045834-081803 (CSS161010; \citealt{coppejans2020aj,Gutierrez2024}), 
AT\,2020xnd \citep{perley2021mnras, ho2022aj, bright2022aj}, 
AT\,2023fhn \citep{Chrimes2024b,chrimes_at2023fhn_2024}, AT\,2024qfm \citep{Fulton2024TNS}, and \AT\, \citep{Pursiainen2025MNRAS, ofek2025}. 
From this small sample of objects, LFBOTs are inferred to possess aspherical ejecta with outflows that are high velocity near the poles ($\sim0.2$c) and lower velocity at the equator (a few $1000\,\rm{km\,s}^{-1}$). 
At early times ($\lesssim15-30$\,d), the ejecta reprocess X-rays produced by the central engine into UV--optical--IR (UVOIR) wavelengths (e.g., \citealt{Piro_Lu_2020,Uno_Maeda_2020,Calderon2021MNRAS.507.1092C,Chen_Shen_2024}). As the ejecta photosphere recedes, H and He features are revealed in their spectra. While such observational features of this class are well-established, the exact ejecta structure and intrinsic nature of LFBOTs is unconstrained owing to the lack of objects with extensive UV--NIR pre- to post-peak photometry and spectroscopy. Only two previous LFBOTs (AT\,2018cow and CSS161010) have extensive, multi-epoch optical spectral sequences. \AT\, presents a rare opportunity to obtain this dataset.

\AT\, is the most luminous LFBOT discovered to date (both in the UV and bolometrically); the first to have pre-peak UV photometry, which led to an extensive UV--optical observational campaign; the second to be sampled in the NIR photometrically and spectroscopically, which revealed the second detected LFBOT NIR excess; the second to be observed with optical polarimetry \citep{Pursiainen2025MNRAS}; and only the third LFBOT with an optical spectral sequence up to 55\,d (rest frame), which revealed unprecedented line profiles of H and He. Despite being more distant (411 Mpc) than the closest known LFBOT AT\,2018cow (60 Mpc), the extreme luminosity of \AT\, ultimately enabled this well-sampled dataset and allowed for the search for short duration optical flares like those observed in AT\,2022tsd \citep{ho2023_2022tsd}, which were not detected \citep{ofek2025}. 
Here in Paper I, we present our multiwavelength observations of \AT\, over the first $\sim100$ days of evolution, with focus on the transient thermal UVOIR emission.  We analyze the broad-band X-ray and radio emission from \AT\, in our companion paper (Nayana et al.~2025, Paper II hereafter). We refer to the results from Paper II where appropriate to build a holistic picture of the event.

This paper is structured as follows. We present our UV, optical and NIR observations (photometry and spectroscopy) in \S\ref{sec:observations} and derive the bolometric luminosity of the transient in \S\ref{Sec:properties}.  
In \S\ref{Sec:featureless}, we discuss the astrophysical implications of the observed featureless spectra maintained for weeks after discovery, and the later emergence of H+He emission lines with unusual profiles. 
Section \S\ref{Sec:NIRexcess} explores the nature of the NIR excess in the context of free-free emission and dust models. We discuss in \S\ref{Sec:Discussion} \AT\, in the context of other LFBOTs and TDEs with similar spectral features, and we conclude in \S\ref{Sec:Conclusions}.

From \cite{Perley24wppredshift}, \AT\, is located at redshift $z=0.0868$, which corresponds to a luminosity distance of  411 Mpc under $\Lambda$CDM cosmological parameters  $H_0=67.4\,\rm{km\,s^{-1}\,Mpc^{-1}}$, $\Omega_m=0.315$, and $\Omega_{\Lambda}=0.685$ \citep{Cosmology20}.
We estimate the time of first light to be MJD =  60578.3. 
We refer to times with respect to this $t_0$, in the observed reference frame, and as UTC unless otherwise stated. Uncertainties in $t_0$ have no impact on our major conclusions. Uncertainties are reported at the 1\,$\sigma$ (Gaussian equivalent) confidence level and upper limits correspond to a 3$\sigma$ statistical level. We correct for  Milky Way (MW) reddening using $R_V = 3.1$ adopting the \cite{F99MWExtCorr} model with $A_V = 0.078$ mag and $E(B-V) = 0.025$ mag. Other than for the \textit{Swift} UV photometry discussed in \S\ref{sec:observations}, we assume host galaxy extinction is negligible and do not apply a correction because of the large transient separation from its host (3.1\arcsec;  \citealt{Perley24wppredshift}) and the lack of narrow absorption lines in the early-time, high-signal-to-noise-ratio (S/N) spectra.

\section{Observations}
\label{sec:observations}

\subsection{UV to NIR Photometry} \label{Photometry}
\begin{figure}[h!]
  \includegraphics[width=1\columnwidth]{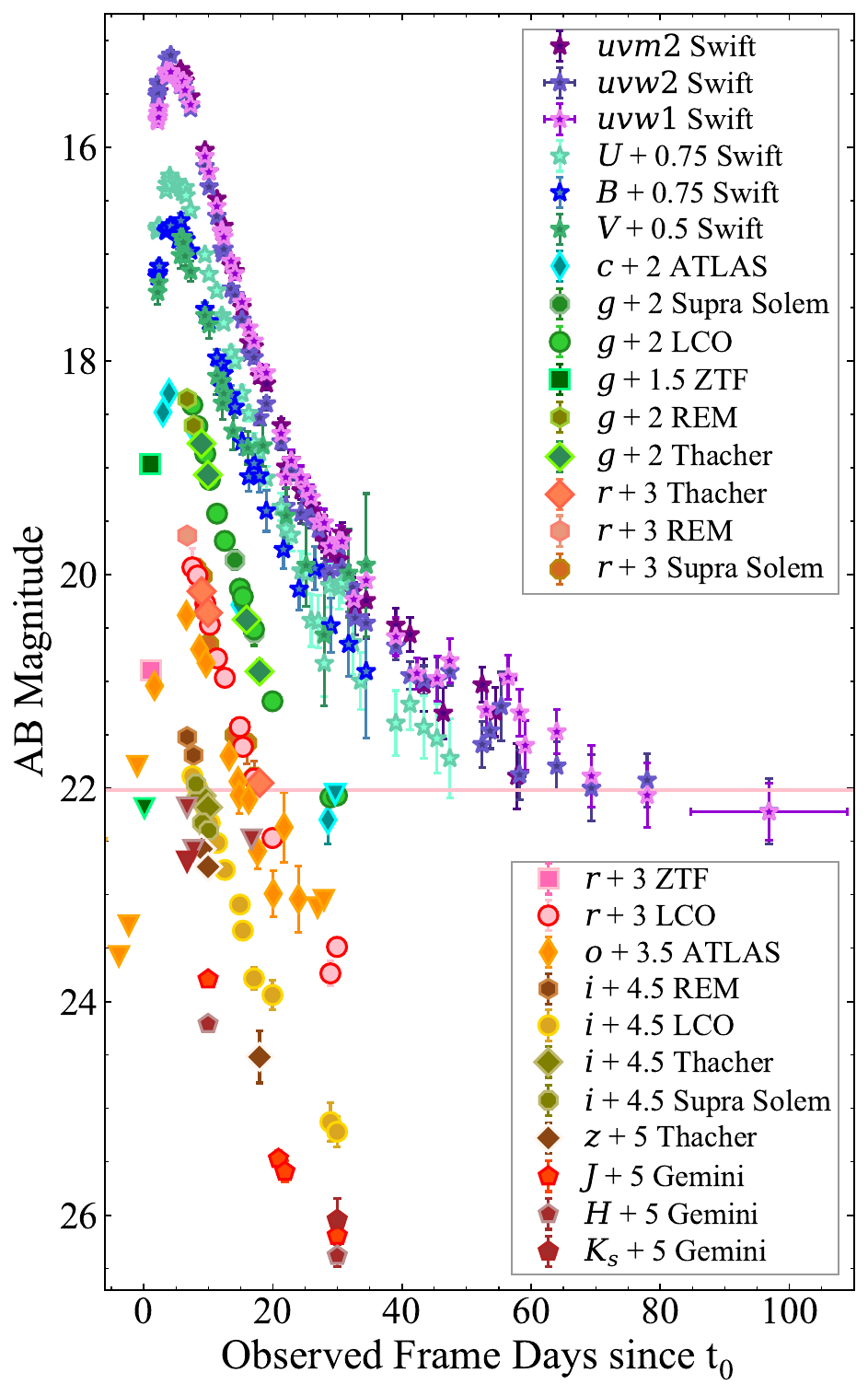}
  \caption{UVOIR light curve of \AT\, corrected for Galactic extinction. Offsets are applied for clarity. Pink horizontal line: pre-transient GALEX NUV emission at the location of the transient, marking the level of UV emission from the host galaxy. At $\gtrsim 80$\,d, the  {Swift} UV photometry is dominated by the host galaxy flux. In \S\ref{Photometry}, we use the GALEX NUV observation to subtract the host galaxy contribution from the {Swift} {$m2$} photometry. Magnitudes are expressed in the AB system \citep{Oke1983ApJ}.} 
  \label{fig:lightcurve}
\end{figure}

Observations with the Ultra-Violet/Optical Telescope (UVOT; \citealt{Roming05}) onboard the Neil Gehrels Swift Observatory \citep{Gehrels04} began on 2024-09-27 at 9:52:34 ($\delta t= 2.1$\,d, PI M. Coughlin). An extensive one-day cadence UVOT campaign was initiated under our Guest Observer program (PI R. Margutti), and covered the period $\delta t =5-59$\,d, followed by a lower-cadence campaign to sample the later-time evolution of \AT\, until $\delta t=119$\,d. This prompt and intense monitoring led to the acquisition of the first UV data during the rise time of an FBOT, and to the most detailed UV data on an FBOT to date.

Ground-based optical and NIR photometry of \AT\, was obtained between 2024 October 2 and 2024 October 25 ($\delta t =7-30$\,d). 
Photometry in filters $g$, $r$, and $i$ was obtained from the Las Cumbres Observatory (LCO) Global Telescope Network 1~m telescopes\footnote{\url{https://lco.global/}} and the Supra Solem Observatory operating in SkiesAway Remote Observatory with a PlaneWave CDK 125 telescope. Additional $g$-, $r$-, and $i$-band photometry as well as three epochs of $z$-band photometry was obtained from the Thacher 0.7\,m telescope in Ojai, CA \citep{Thacher}.
Three epochs of optical and NIR observations in filters $g$, $r$, $i$, $J$, and $H$ (epoch 1), $g$, $r$, $i$, and $J$ (epoch 2), and $J$ (epoch 3) were obtained by the 0.6 m robotic Rapid Eye Mount telescope (REM; \citealt{Zerbi01,Covino04}), located at the European Southern Observatory (ESO) at La Silla (Chile). A fading source was detected consistent with the transient position in all REM optical observations; however, no infrared counterpart was detected in any of the three REM NIR observations. 
An additional four epochs of NIR imaging were obtained with the Flamingos-2 instrument \citep{Eikenberry2004_F2,Eikenberry2012_F2} mounted on the Gemini-South Telescope. 
The first three Gemini epochs are derived from the acquisition images for NIR spectroscopy in the $J$ and $H$ bands; the fourth epoch was observed in $J$, $H$, and $Ks$.
The source was detected at all Gemini epochs. 

We also collected publicly available photometry from the Transient Name Server\footnote{\url{https://www.wis-tns.org/object/2024wpp}} (TNS) AstroNotes and the Asteroid Terrestrial-impact Last Alert System (ATLAS; \citealt{Tonry2018PASP,Smith2020PASP}). ATLAS data were obtained through the ATLAS forced-photometry server\footnote{\url{https://fallingstar-data.com/forcedphot/}}. ATLAS photometry is listed in Table \ref{tab:atlas_phot_all} and public ZTF photometry reported to TNS \citep{Ho2024TNSAN.272....1H} is listed in Table \ref{tab:ZTF_public_phot}. Magnitudes in these tables are corrected for MW extinction.

All data were reduced using standard procedures. 
{Swift}-UVOT observations
span the wavelength range $\lambda_c=1928$ \AA\, ($w2$ filter) -- $\lambda_c=5468$ 
\AA\, ($V$ filter; central wavelengths listed). We extracted the UVOT photometry following standard practice and updated zero-points (e.g., \citealt{Brown09}). Specifically, we used a 5\arcsec-radius source region centered at the location of \AT\, and a 35\arcsec-radius source-free region to estimate the background contribution. We merged individual exposures to reach a minimum S/N $\gtrsim 10$.
We estimate the host galaxy flux contribution to be negligible at early times ($\delta t<45$\,d) for all filters. 
Final observations in filters $w1$ and $w2$ near $100$\,d are assumed to be host dominated and we use these measurements to host-correct the $w1$ and $w2$ data. Pre-explosion observations from the Galaxy Evolution Explorer ({GALEX}; \citealt{Martin2005ApJGALEX}) measured the host galaxy UV emission to be $m_{NUV} = 22.00 \pm 0.48$ mag. The {GALEX-NUV} to {Swift-{$m2$}} filter correction based on the best-fitting host galaxy spectral energy distribution (SED) from \textit{BLAST}\footnote{\url{https://blast.scimma.org/transients/2024wpp/}} is $\delta_{mag}\approx 0.016$ mag. We thus use the filter-corrected {GALEX-NUV} measurement (with added 5\% systematic uncertainty to account for filter transmission differences) to correct for the host contribution to the $m2$ {Swift} observations.
{Swift}-UVOT photometry (with correction for MW extinction but without host correction) is listed in Tables \ref{tab:swift_phot_UBV} (optical filters) and \ref{tab:swift_phot_UVW} (UV filters).

Supra Solem images were flat-fielded, bias-corrected, and dark-corrected using Maxim DL processes\footnote{\url{https://cdn.diffractionlimited.com/help/maximdl/MaxIm-DL.htm}}, implemented automatically by the ACP Observatory Control Software\footnote{\url{http://scheduler.dc3.com/}}. Template subtraction was performed on the science images with the High Order Transform of Psf ANd Template Subtraction (HOTPANTS; \citealt{Becker15}) code with pre-explosion Pan-STARRS images. Aperture photometry was performed on the subtracted images using the \texttt{photutils} package in \texttt{astropy} \citep{2013A&A...558A..33A}, and flux-calibrated from zero-points derived from Pan-STARRS DR2 sources \citep{Flewelling18}. Supra Solem photometry (MW extinction corrected) is listed in Table \ref{tab:supra_solem_phot_all}.

For the LCO photometry, calibrated BANZAI frames were downloaded from the LCO archive and aperture photometry was performed using the procedure outlined above for Supra Solem Observatory data. LCO photometry (MW extinction corrected) is listed in Table \ref{tab:LCO_phot}.

REM data reduction was performed with the REM reduction pipeline. After bias subtraction, nonuniformities were corrected using a normalized flat-field frame processed with tools from the Swift Reduction Package (SRP).\footnote{ \url{http://www.me.oa-brera.inaf.it/utenti/covino/usermanual.html}} NIR data were sky-subtracted using the median of individual frames. Frame registration was performed using the Python-based software Astroalign \citep{Beroiz20}, and astrometric solutions were derived against Gaia DR3 stars \citep{Vallenari23}.
Aperture photometry was performed, calibrating against Pan-STARRS DR2 sources (PS1; \citealt{Flewelling18}).
Upper limits on the NIR photometry were derived using calibration against 2MASS stars (2MASS; \citealt{Skrutskie06}).
REM photometry (MW extinction corrected) is listed in Table \ref{tab:REM_phot_all}.

Flamingos-2 
images were reduced with standard recipes in DRAGONS \citep{DRAGONS,DRAGONS:zendo}
Aperture photometry was performed with \texttt{photutils} and flux calibration was done using zero-points derived from 2MASS stars. Gemini photometry (MW extinction corrected) is listed in Table \ref{tab:gemini_phot}.  

Thacher $griz$ images were processed using standard reduction procedures using {\tt photpipe} \citep{Rest05}.  All images were calibrated using flat-field and bias frames from the same night and instrumental configuration, and astrometrically calibrated to Gaia DR3 calibrators.  We transformed each image to a regular image coordinate frame with {\tt SWarp} \citep{swarp}, and then performed point-spread-function (PSF) photometry using {\tt DoPhot} \citep{Schechter93}.  Finally, we photometrically calibrated photometry from each field using Pan-STARRS DR2 sources. Thacher photometry (MW extinction corrected) is listed in Table \ref{tab:thatcher_phot_all}.
 
All photometric observations are presented in Fig.~\ref{fig:lightcurve}.

\subsection{Optical and NIR Spectroscopy}\label{Subsec:spectroscopy}
To estimate our $t_0$, we linearly extrapolate the rise rate in flux between the first ZTF $g$-band marginal detection and the subsequent detection \citep{Ho2024TNSAN.272....1H}. 

\begin{figure*}[ht!]
  \centering
  \includegraphics[width=2.1\columnwidth]{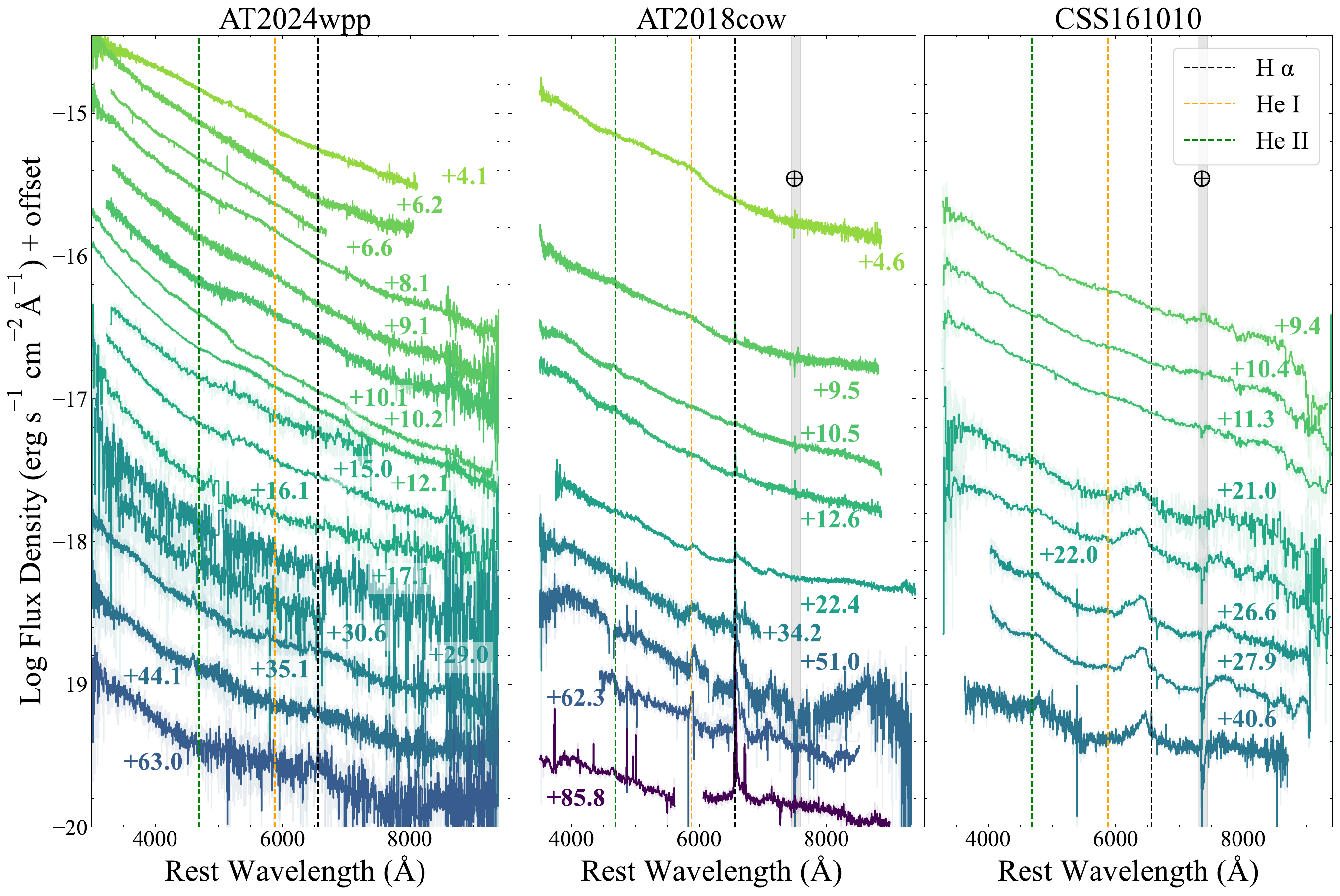}
  \caption{\emph{Left:} Collated AT\,2024wpp optical spectral series spanning +4.1 to +63.0 days from $t_0$ (observed frame). Spectra are plotted in the rest frame and are binned for visual clarity. For comparison, coeval optical spectra of the only other well-sampled LFBOTs, AT\,2018cow \citep{Margutti19} and CSS161010 \citep{Gutierrez2024}, are plotted in the middle and right panels respectively and identified by the observed-frame epoch of observation. We note that the +4.1 d \AT\, spectrum has an unreliable spectral slope. 
  }
  \label{fig:optical_spec}
\end{figure*}

\begin{figure*}[ht!]
  \centering
  \includegraphics[width=1.25\columnwidth]{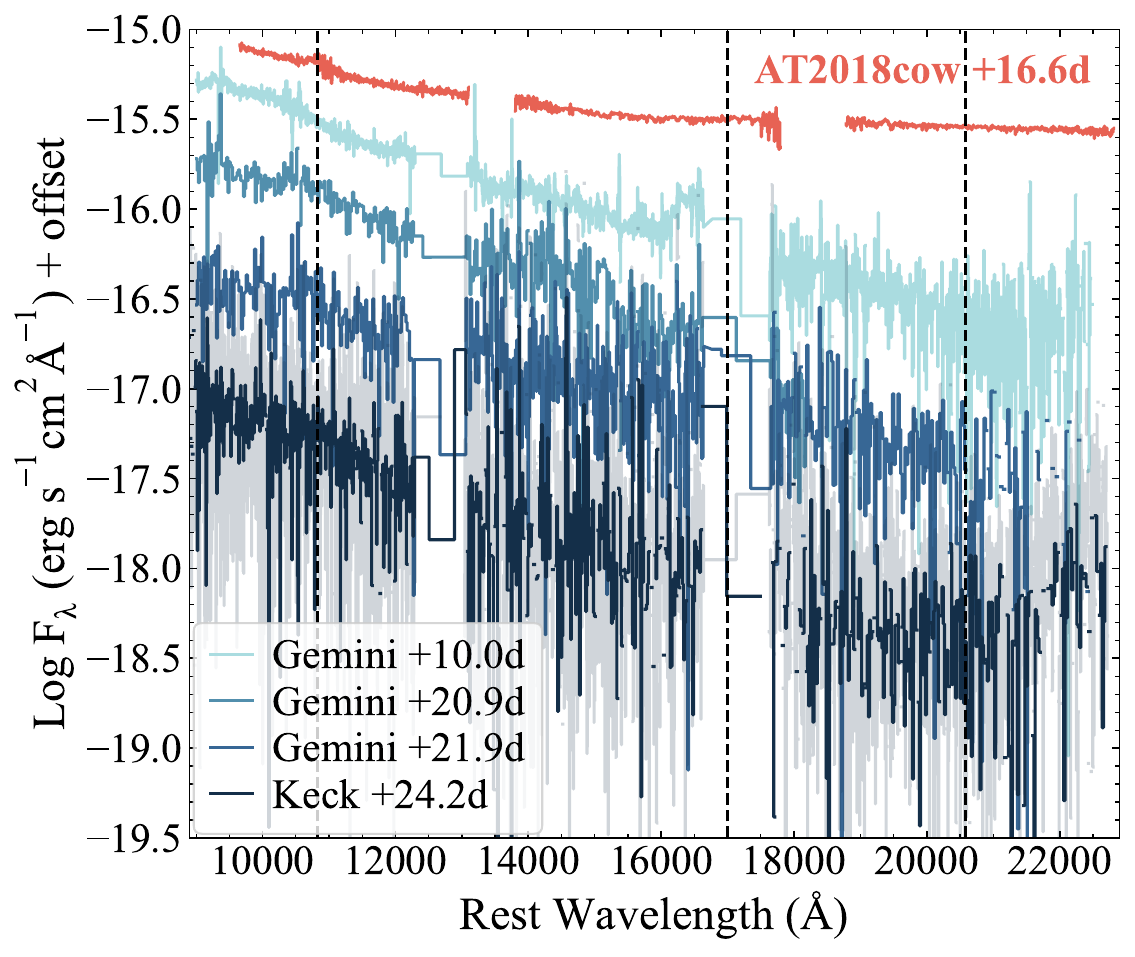}
  \includegraphics[width=0.74\columnwidth]{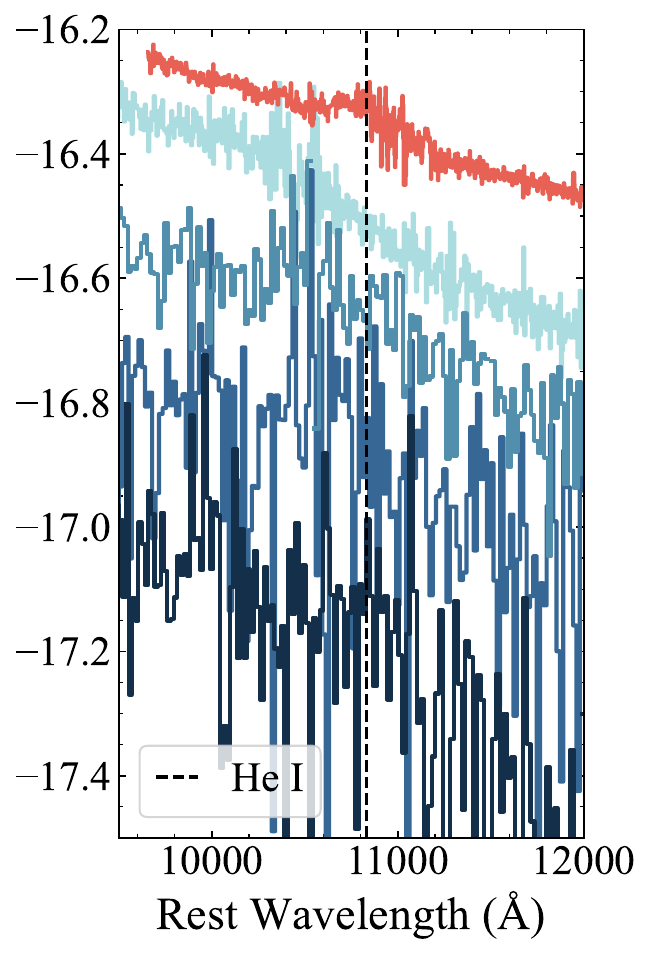}
  \caption{\emph{Left:} AT\,2024wpp NIR spectral series from +10.0\,d to +24.2\,d after $t_0$. Spectra are plotted in the rest frame and are binned and scaled for clarity. For comparison, a +16.6\,d spectrum of AT\,2018cow \citep{Margutti19}, the only other LFBOT with NIR spectra, is plotted. In contrast to the AT\,2018cow spectrum, the AT\,2024wpp spectra show no signs of He feature development by day 24. \emph{Right:} Zoom-in on the 1.08\,$\mu$m He I feature.  }
  \label{fig:NIR_spec}
\end{figure*}

We present 16 optical spectra of \AT\, observed  4--63 d from explosion in Fig. \ref{fig:optical_spec}. We list these spectra in Table \ref{tab:optical_wpp_spec} and overview reduction details below. 

We obtained seven optical spectra with the Kast double spectrograph \citep{miller1994} mounted on the Shane 3~m telescope with a 2.0\arcsec-wide slit. Three of these Kast spectra were reduced using the {\tt UCSC Spectral Pipeline}\footnote{\url{https://github.com/msiebert1/UCSC\_spectral\_pipeline}} \citep{Siebert20}, a custom data-reduction pipeline based on procedures outlined by \citet{Foley03}, \citet{Silverman2012}, and references therein, while the other Kast spectra were reduced in an equivalent manner.  The two-dimensional (2D) spectra were bias-corrected, flat-field corrected, adjusted for varying gains across different chips and amplifiers, and trimmed. One-dimensional spectra were extracted using the optimal algorithm \citep{Horne1986PASP}.  The spectra were wavelength-calibrated using internal comparison-lamp spectra with linear shifts applied by cross-correlating the observed night-sky lines in each spectrum to a master night-sky spectrum.  Flux calibration and telluric correction were performed using standard stars at a similar airmass to that of the science exposures.  We combine the sides by scaling one spectrum to match the flux of the other in the overlap region and use their error spectra to correctly weight the spectra when combining.  More details of this process are discussed elsewhere \citep{Foley03, Silverman2012, Siebert20}. The $+$4.1~d Kast spectrum was obtained at airmass 1.7 with the slit oriented 20--30~deg away from the parallactic angle \citep{Filippenko1982PASP}, so this spectrum experienced enhanced loss of blue light, making the blue spectral slope unreliable.

We also obtained five optical spectra with LRIS \citep{Oke1995PASP..107..375O}  
mounted on the Keck I 10~m telescope. 
LRIS spectra were reduced and calibrated similarly to the Kast spectra.
Low-order polynomial fits to calibration-lamp spectra were
used to establish the wavelength scale, and small adjustments
derived from night-sky lines in the object frames were applied. The +12.1 d LRIS spectrum was reduced in an equivalent manner with LPipe \citep{Perley2019PASP}. 
LRIS is equipped with an atmospheric dispersion corrector, thereby precluding differential slit losses.

We obtained two optical spectra of AT 2024wpp with the Southern African Large Telescope (SALT) Robert Stobie Spectrograph \citep[RSS;][]{Smith_RSS_2006}. The first observation was taken on 2024 Oct. 10, and the second on 2024 Oct. 25. We used a $1.5\arcsec$-wide slit and the PG0900 grating in two tilt positions to cover the blue part of the spectrum without detector chip gaps. The data were reduced using RUSALT, a custom pipeline based on PySALT \citep{Crawford_pysalt_2010} which uses standard Pyraf \citep{pyraf_2012} spectral reduction routines such as wavelength and relative flux calibration, 1D extraction, and the removal of cosmic rays and telluric absorption.

We obtained an optical spectrum of \AT\, using the DeVeny optical spectrograph mounted on the 4.3~m Lowell Discovery Telescope (LDT) on 2024-10-10 (PI E. Hammerstein). 
The spectrum was reduced using \texttt{PypeIt} \citep{Prochaska2020_pypeit,Prochaska_pypeit_zenodo} and standard optical spectroscopic reduction techniques, including bias subtraction, flat-fielding, flux calibration, coaddition, and telluric correction.
Optical spectra are presented in Fig.~\ref{fig:optical_spec} and Table \ref{tab:optical_wpp_spec}.

We present a total of four epochs of NIR spectroscopy of \AT\, in Fig. \ref{fig:NIR_spec}. These spectra are listed in Table \ref{tab:nir_wpp_spec} and we overview reduction details below. 

Three epochs of $JH$ and two epochs of $HK$ NIR spectra of \AT\, were obtained with the Flamingos-2 instrument \citep{Eikenberry2004_F2,Eikenberry2012_F2} mounted on the Gemini-South Telescope (PI N. LeBaron). The additional $JH$ spectrum was obtained due to technical issues preventing the full $JH+HK$ spectroscopy sequence from being obtained on 2024 October 16. 
The spectra were reduced with 
\texttt{PypeIt}, which performed flat-fielding, background subtraction, and source detection and extraction. The science spectra were then flux-calibrated, coadded, and corrected for telluric absorption, using the A0\,V star HIP12858 which was observed directly after \AT{}. 

We also observed \AT\, with the Near-InfraRed Echellette Spectrometer (\citealt{Wilson2004SPIE.5492.1295W}; NIRES) on the Keck II 10 m telescope on 2024 October 20 as part of the Keck Infrared Transient Survey (KITS; \citealt{tinyanont2024pasp}). The observations were performed with two sets of the ABBA dithering pattern to sample the sky background, with a total exposure time of 2200\,s. The A0\,V star HIP14627 was observed immediately after the LFBOT to provide flux and telluric calibration. We reduced the data using \texttt{PypeIt} following the procedure outlined by \citet{tinyanont2024pasp}. 

\begin{figure*}[ht!]
  \centering
    \includegraphics[width=\columnwidth]{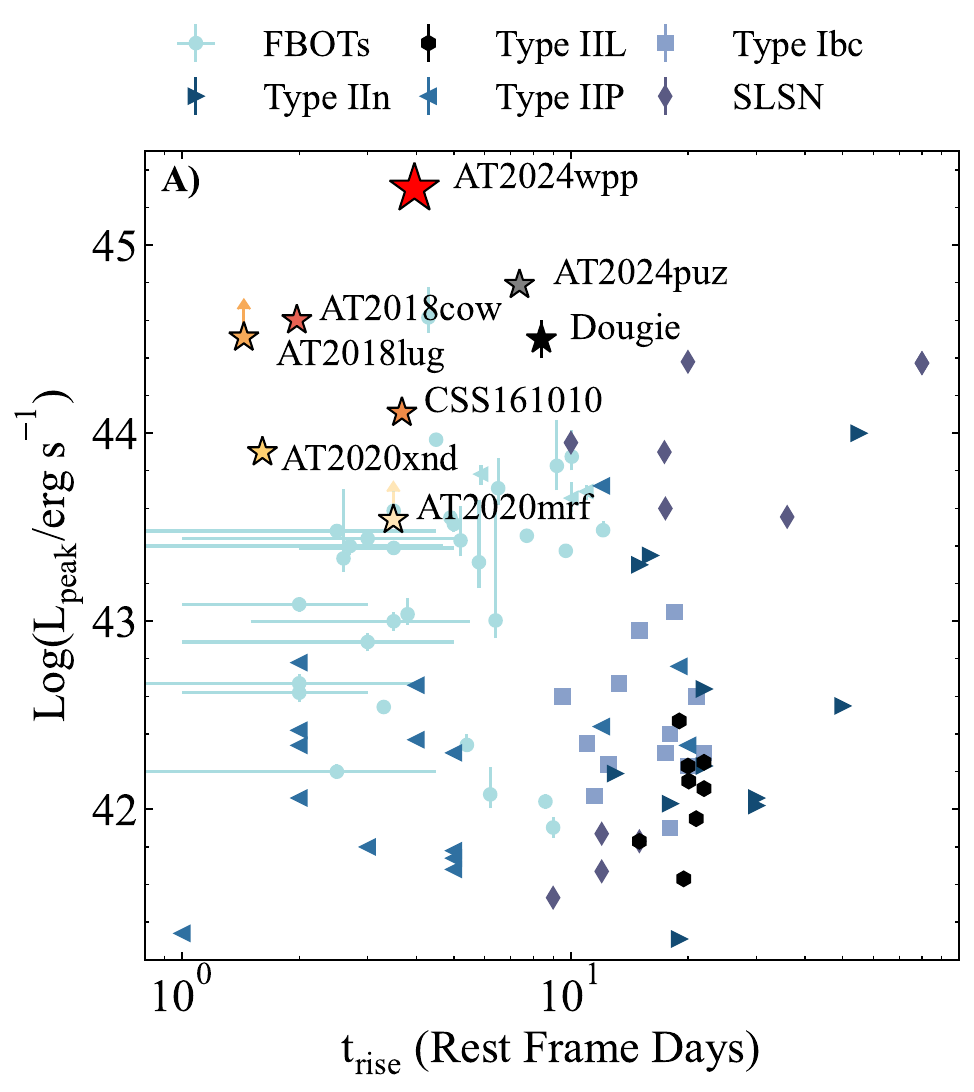}
\includegraphics[width=\columnwidth]{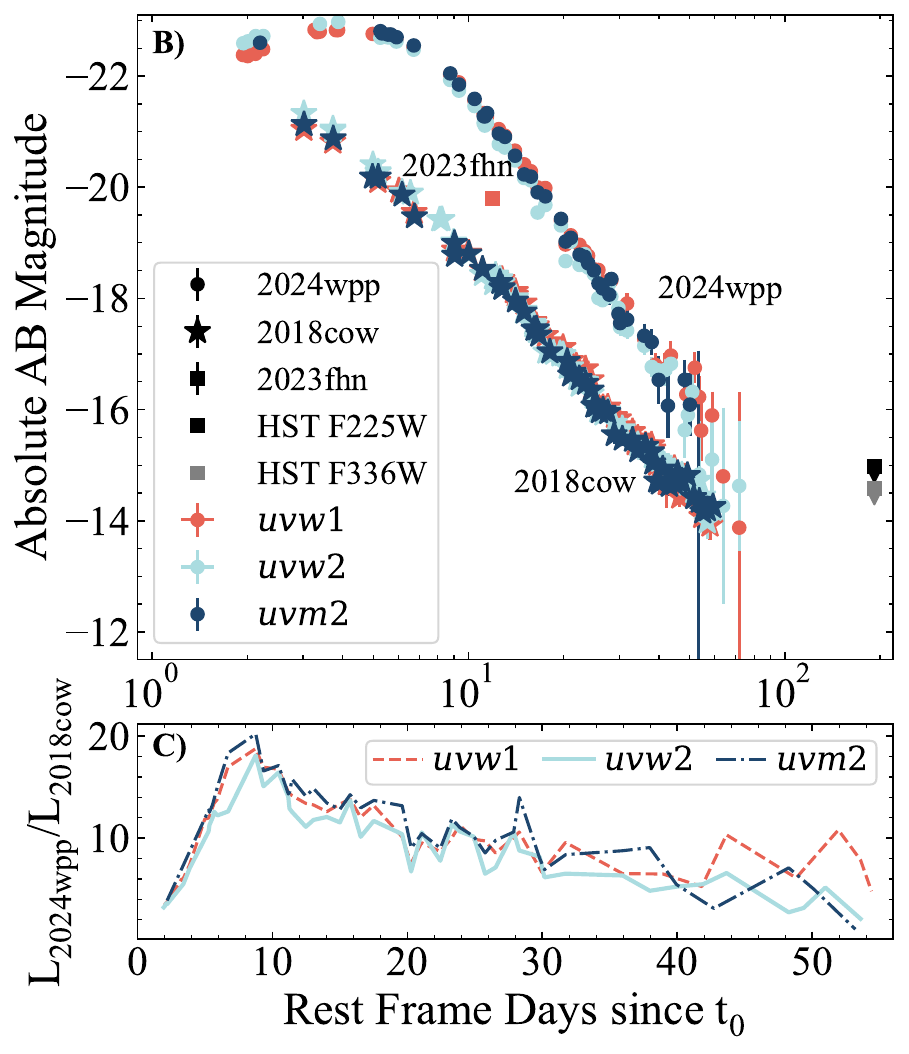}
 
  \caption{AT\,2024wpp is the most luminous LFBOT discovered to date. \emph{Panel A:} AT\,2024wpp in the phase space of peak bolometric luminosity vs. rise time (rest frame) for FBOTs and other transients. Colored (grayscale) star symbols represent LFBOTs (likely LFBOTs). References: AT\,2018cow \citep{Margutti19,perley2019mnras}, CSS\,161010 \citep{Gutierrez2024}, AT\,2018lug \citep{ho2020aja}, AT\,2020xnd \citep{perley2021mnras}, AT\,2020mrf \citep{yao2022aj}, Dougie \citep{Vinko2015ApJ}, AT\,2024puz \citep{Somalwar2025}, FBOTs \citep{drout2014aj, Pursiainen2018MNRAS, Arcavi2016ApJ}, Type Ibc SNe \citep{Taubenberger2006MNRAS, patat_metamorphosis_2001, valenti2008mnras, Ferrero2006A&A, Richmond1996AJ,Valenti2011MNRAS, Modjaz2009ApJ,hunter2009aa,Yoshii2003ApJ,Mazzali2000ApJ,Mazzali2006ApJ,Stritzinger2002AJ,Pignata2011ApJ}, SLSNe \citep{quimby_hydrogen-poor_2011, inserra2013a}, Type IIP SNe \citep{Hamuy_2003}, Type IIn SNe \citep{Kiewe2012ApJ...744...10K, Margutti14}, Type IIL SNe \citep{Arcavi2012ApJ}. We calculate peak pseudobolometric luminosities of LFBOTs AT\,2018lug, AT2020xnd, and AT2020mrf using their peak $g$-band magnitudes (i.e., $L_{\rm{pk}} = \nu_g L_{\rm{pk},\nu_g}$). For FBOTs, SLSNe, and some SNe~Ib/c, we similarly plot pseudobolometric luminosities.
  \emph{Panel B:} Comparison of the three best-sampled LFBOTs in the UV (extinction-corrected, host-subtracted  absolute magnitudes). For AT\,2023fhn, we report the  HST results from \citet{Chrimes2024b} for aperture photometry performed using a $0.4''$ annulus background and an extinction-corrected {Swift} {$w1$} observation (see Appendix \S\ref{AT2023fhn_appendix} for reduction details).
  {Swift}-UVOT observations of \AT\, captured the rise of a LFBOT and the evolution of a LFBOT at $\delta t\gtrsim 60$\,d for the first time at UV wavelengths. \emph{Panel C:} \AT\, is  $\sim 4.5$ times more UV luminous than the prototypical event AT\,2018cow at peak, and it is the most UV-luminous FBOT discovered to date.}
  \label{fig:UVcombined}
\end{figure*}

\begin{figure}[ht!]
  \includegraphics[width=.9\columnwidth]{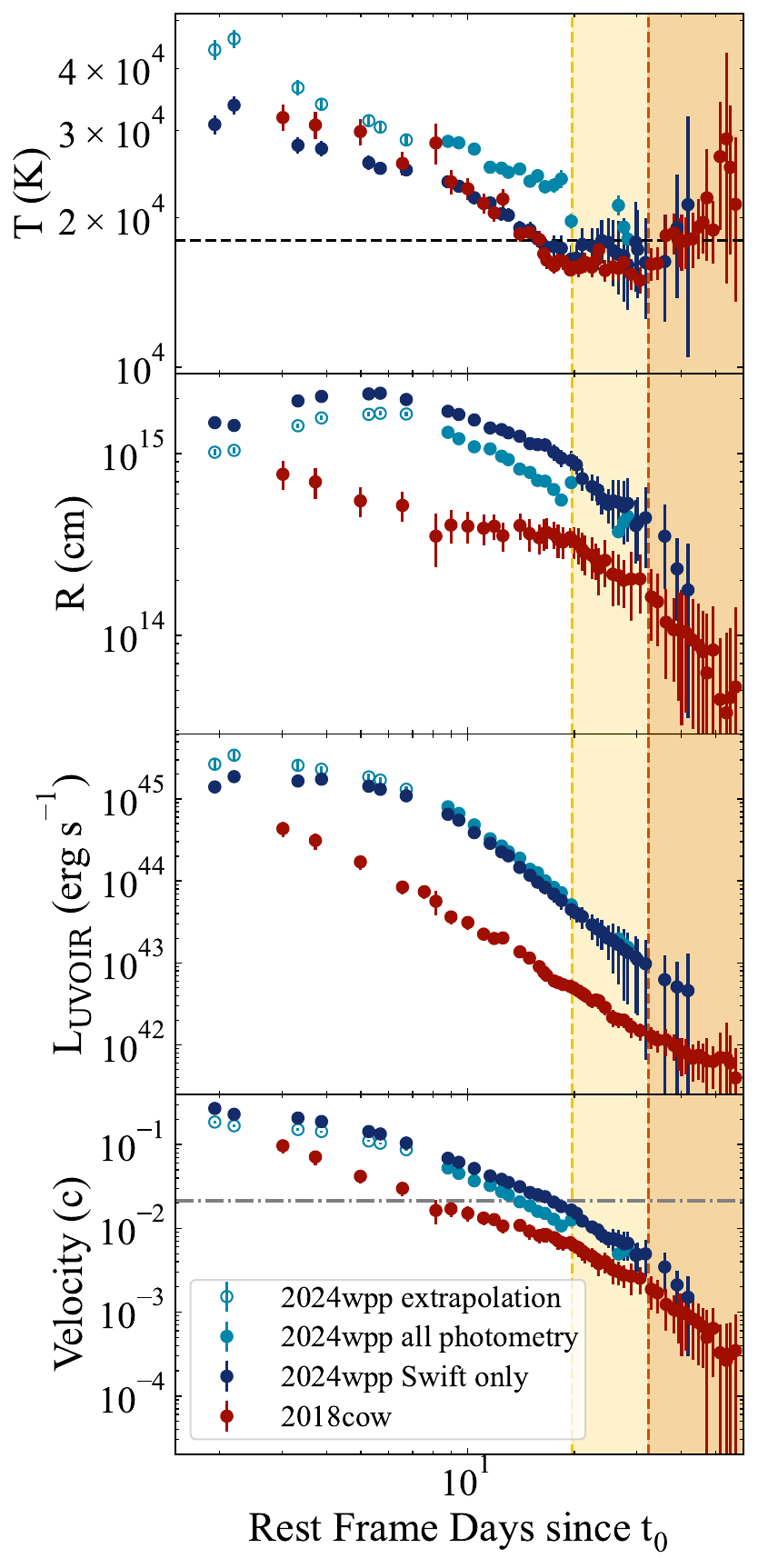}
  \vskip -0.4 cm
  \caption{Temporal evolution of the best-fitting blackbody temperature $T$, radius $R$, bolometric luminosity $L_{\rm{UVOIR}}$, and blackbody expansion velocity of \AT\, (shades of blue)  compared to AT\,2018cow (red) from \cite{Margutti19}. Dark (light) blue filled points: results from the Swift (Swift+LCO+Gemini) photometry.   
  Open circles assume the color extrapolation presented in Fig.~\ref{fig:color} for {$w1$} -- \textit{r} and {$w1$} -- \textit{i} between days +2 and +7 from $t_0$. %
  AT\,2024wpp reaches $T>30{,}000$\,K (potentially as high as $\gtrsim 40{,}000$\,K), and similarly to AT\,2018cow, the $T$ plateaus at late times around $T\approx 20{,}000$\,K (horizontal, dashed black line in the upper panel).  
  Yellow and orange shaded areas indicate the time of emergence of spectroscopic features in AT\,2018cow and AT\,2024wpp, respectively. 
  The horizontal, gray dashed-dotted line marks a velocity of $\sim6400$ km s$^{-1}$, which corresponds to the observed blueshift velocity of spectroscopic features that emerge between +16\,d and +30\,d (see Fig. \ref{fig:optical_spec}).
  }
  \label{fig:BB_fits}
\end{figure}

\begin{figure*}[ht!]
\centering
  \includegraphics[width=2.1\columnwidth]{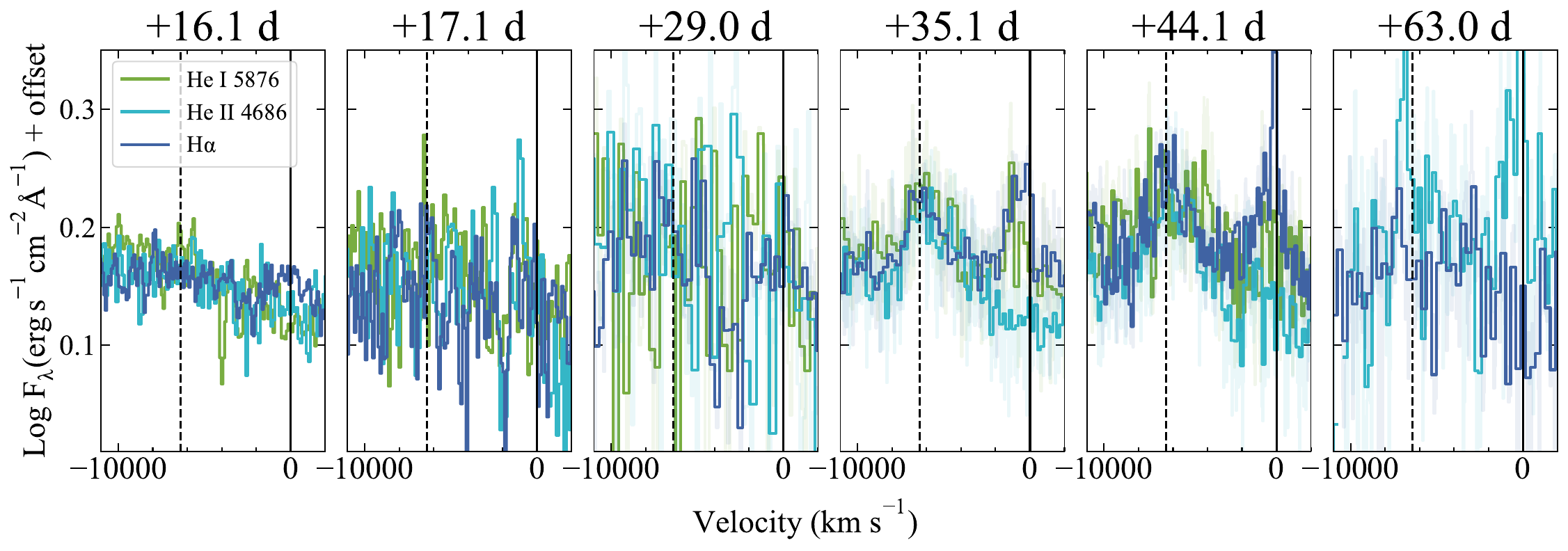}
  \caption{By +35 d, \AT\, shows the emergence of clear H$\alpha$ and He features blueshifted by $\sim6400\,\rm{km\,s}^{-1}$(dashed black line). The FWHM values of the blueshifted features stay consistently near $\sim2000\,\rm{km\,s}^{-1}$ over the final three epochs. A blend of narrow host galaxy emission and a slightly broader spectral feature base from \AT\, is also evident at $0\,\rm{km\,s}^{-1}$, especially for H$\alpha$ and He I. We do not identify a He I feature in the +63 d spectrum, and thus omit that wavelength region for visual clarity. 
  }
  \label{fig:optical_lines}
\end{figure*}

\begin{figure}[ht!]
\hskip -0.5 cm \includegraphics[width=1\columnwidth]{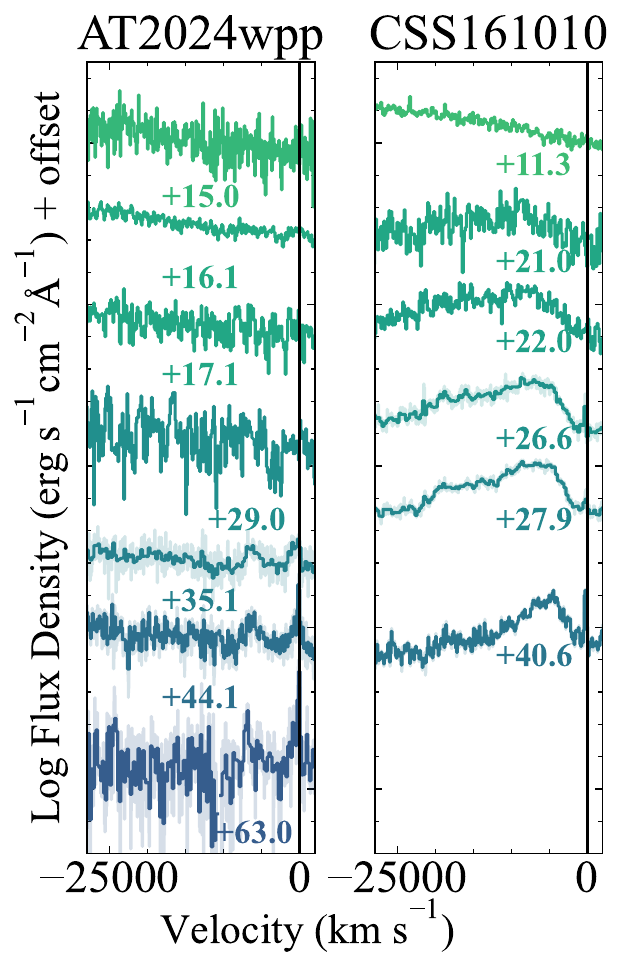}
 
  \caption{Evolution of the H$\alpha$ line in \AT\, (left) and CSS161010 (right; \citealt{Gutierrez2024}), the two LFBOTs with evidence for blueshifted emission. The $v=0\,\rm{km\,s}^{-1}$ component in CSS161010's spectrum is narrow host galaxy emission. In \AT, the component is broader ($\rm{FWHM}\approx 2000\,\rm{km\,s}^{-1}$, similar to the blueshifted component; see Fig.~\ref{fig:Line_params}); thus, we identify it as a transient emission feature.}
  \label{fig:optical_halpha}
\end{figure}

\section{Bolometric Luminosity and inferred properties} \label{Sec:properties}
Similar to AT\,2018cow, at $\delta t\le45$\,d the UV to optical radiation from \AT\, is dominated by a blackbody spectrum. The extremely blue colors and color evolution of \AT\, impose non-negligible deviations from the standard UVOT count-to-flux conversion factors. We account for this effect self-consistently in our blackbody fits by following the prescriptions by \citet{Brown10} --- that is, we iteratively recalibrate the fluxes with the input blackbody temperature until the input and output temperature agree to within uncertainties, as done for AT\,2018cow \citep{Margutti19}. We find that with a peak bolometric luminosity $L_{\rm{pk}}\approx 10^{45}\,\rm{erg\,s^{-1}}$ and a rise time of $t_{\rm{rise}}\approx4$\,d (estimates in agreement with \citealt{Pursiainen2025MNRAS}), AT\,2024wpp is the most luminous known FBOT  (Fig.~\ref{fig:UVcombined}).  Reaching an absolute UV magnitude of $\sim -22.98$, AT\,2024wpp was $\sim 4.5$ times more UV-luminous at peak than the prototypical event AT\,2018cow.  

We show the best-fitting blackbody parameters (temperature $T(t)$ and radius $R(t)$) in Fig. \ref{fig:BB_fits}. AT\,2024wpp displays a high temperature ($T>30{,}000$\,K) in the first week of evolution and maintains $T\gtrsim 20{,}000$\,K until the end of our monitoring, again in strict similarity with AT\,2018cow, but in stark contrast with SNe (see, e.g., \citealt{Dessart2011MNRAS.414.2985D}). Our inferred temperatures are higher around peak than those presented by \citet{Pursiainen2025MNRAS} which could be attributed to our iterative recalibration of the UV fluxes as described previously. Without this correction, we obtained similar temperatures to \citet{Pursiainen2025MNRAS} and our fits settle at $T\gtrsim 20{,}000$\,K consistent with their analysis at epochs at which the recalibration is less impactful. We note that a ``lack of cooling with time'' is also a hallmark observational feature of TDEs \citep{vanVelzen2011ApJ...741...73V}. The initial blackbody radius of \AT\, is $R\approx (1-2)\times 10^{15}\,\rm{cm}$. This radius shows limited increase in the first few days before later decreasing monotonically with time, again in contrast with ordinary SNe where the photospheric radius typically shows a linear increase with time during the first few weeks \citep{Dessart2011MNRAS.414.2985D}. This evolution implies an initial average blackbody ``expansion velocity'' (defined as $R/t$) as high as  $\sim (0.2-0.3)\,c$, decreasing to $< 6400\,\rm{km\,s^{-1}}$ at $\delta t\ge20$\,d. We note that the initial high expansion velocities are similar to those inferred from radio modeling of the blast wave in Paper II, by analogy to AT\,2018cow. This, in addition to the more luminous \AT\, having slightly higher velocities than AT\,2018cow, points to a connection between LFBOT optical and radio emission components. We also note that blueshifted spectral features with $v\approx 6400\,\rm{km\,s^{-1}}$ appear in the time period $\delta t=16-30$\,d (Fig. \ref{fig:optical_lines}), consistent with the idea that the recession of the blackbody radius inward revealed slower material in the LFBOT. The velocity and time of appearance of the spectral features associated with slowly moving material makes it consistent with ejecta launched at $t_0$. 

While the overall UV-to-NIR bolometric emission is well fit by a blackbody continuum, we find evidence for an excess of NIR emission at $\delta t=30.0$\,d, but no evidence for a NIR excess from our broad-band photometry at day 10.3. \S\ref{Sec:NIRexcess} discusses the observational properties of the NIR excess, its connections with a similar excess reported for AT\,2018cow, and potential scenarios that can explain our observations.

We end with a few considerations. First, we derive an order-of-magnitude estimate of the ejecta mass  $M_{\rm{ej}}$ at peak brightness, under the assumption that the rise time $t_{\rm{rise}}$ reflects the diffusion time of radiation from a centrally located source within ejecta expanding with typical velocity $v_{\rm{ej}}$. Following, for example, \citet{Margutti19},
\begin{equation}
    M_{\rm{ej}} \approx \frac{4\pi t_{\rm{rise}}^2 v_{\rm{ej}} c}{\kappa} \approx 2.0\, \rm{M_\odot} \left( \frac{0.1 \text{ cm}^2 \text{g}^{-1}}{\kappa} \right) \left( \frac{v_{\rm{ej}}}{0.3c} \right) \left( \frac{t_{\rm{rise}}}{4 \text{ d}} \right)^2\, ,
    \label{eqn:ejectamass}
\end{equation}
where $\kappa$ is an order-of-magnitude estimate of the opacity,\footnote{The electron-scattering opacity for fully ionized, H-depleted ejecta is $\kappa_{\rm{es}}\approx 0.2$\,cm$^{-2}$\,g$^{-1}$.} and we have adopted an ejecta velocity of $0.3$\,c as indicated by our blackbody fits in Fig. \ref{fig:BB_fits} (for  $v_{\rm{ej}}=0.2$\,c, $M_{\rm{ej}}\approx1.4\,\rm{M_{\sun}}$). The implied corresponding kinetic energy of the optically emitting material is large: $E_{\rm{k}}\approx (5-15)\times 10^{52}$\,erg (compared to $ (0.03-0.3)\times 10^{52}$\,erg inferred for AT\,2018cow; \citealt{Margutti19}), effectively ruling out ordinary stellar explosions.
The estimated $M_{\rm{ej}}$ for \AT\, is larger than that inferred for AT\,2018cow  with the same approach \citep[$\sim 0.1-0.5\,\rm{M_\odot}$ from][]{Margutti19}, consistent with the longer rise time to peak, but overall similar blackbody initial expansion velocity. We note that Eq.~\ref{eqn:ejectamass} is an upper limit as $t_{\rm{rise}} = \max(t_{\rm{diff}},t_{\rm{visc}})$ (see example in \citealt{metzger2022aj}). 
Under other methods and assumptions (e.g.,~following \citealt{Roth16, Matsumoto2021MNRAS.502.3385M}), we find $M_{\rm ej}\lesssim 1\,\rm{M_{\odot}}$.

\begin{figure}[h!]
\centering
  \includegraphics[width=\columnwidth]{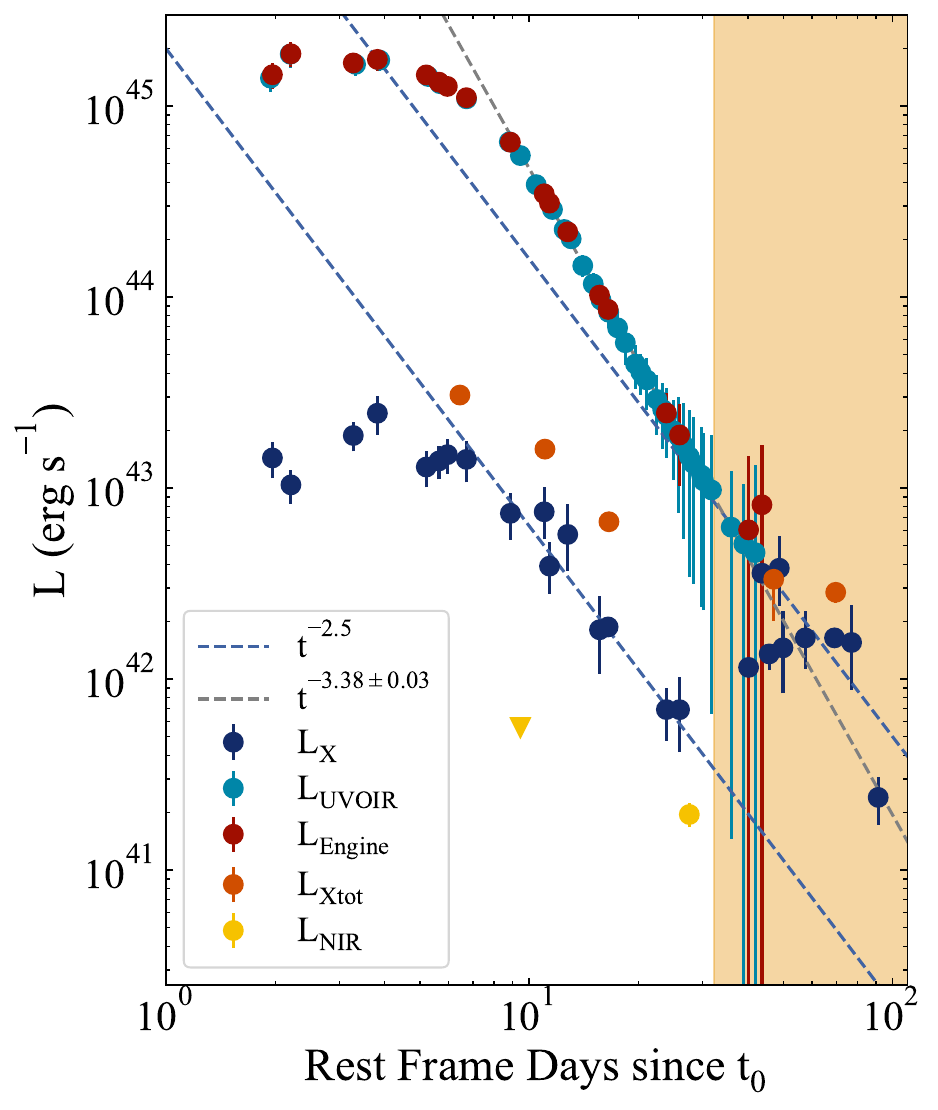}
  \caption{Soft X-ray (0.3--10 keV; $L_{\rm{X}}$, dark-blue circles), broad-band X-ray (0.3--30 keV; $L_{\rm{Xtot}}$, orange circles), NIR (fitted power-law luminosity in the range 2000--24,000\,\AA; $L_{\rm{NIR}}$, yellow circles), and  UVOIR bolometric luminosity ($L_{\rm{UVOIR}}$, teal circles) evolution  of \AT{}\, in its first 100 days.  We also show the ``engine luminosity'' $L_{\rm{engine}}\equiv L_{\rm{X}}+L_{\rm{UVOIR}}$ with dark-red circles. We find that at $10<\delta t <30$\,d, $L_{\rm{engine}}\propto t^{-3.4}$ (gray dashed line) and that a similar scaling applies to  $L_{\rm{UVOIR}}$ at $10\,\rm{d}< \delta t<30\,\rm{d}$, while $L_{\rm{X}}\propto t^{-2.5}$ (blue dashed line). At $\delta t>30\,$d, there is some indication that the decay of $L_{\rm{engine}}$ flattens slightly toward the $t^{-2.5}$ power law. The orange shaded area marks the time of appearance of clear  spectral features ($\delta t \approx16-30$\,d, Fig. \ref{fig:optical_spec} and \ref{fig:optical_lines}).  X-ray data from Paper\,II. 
  }
  \label{fig:Lengine}
\end{figure}

Post peak, the UVOIR bolometric light curve decays as $L_{\rm{UVOIR}}\propto t^{-3.4}$, steeper than the evolution of AT\,2018cow \citep{perley2019mnras,Margutti19} for which $L_{\rm{UVOIR}}\propto t^{-2.5}$ (Fig. \ref{fig:BB_fits}). Defining the ``engine luminosity'' as $L_{\rm{engine}}\equiv L_{\rm{X}}+L_{\rm{UVOIR}}$ (where $L_{\rm{X}}$ is the soft X-ray luminosity integrated in the range 0.3--10 keV; $L_{\rm{engine}}$ is a relevant quantity \emph{if} the  thermal UVOIR emission and the soft X-rays are manifestation of the same physical component; see \citealt{Margutti19}), and using the X-ray results from Paper II, we find a similar temporal evolution  $L_{\rm{engine}}\propto t^{-3}$ at $\delta t\le 100$\,d, again steeper than in AT\,2018cow, for which $L_{\rm{engine}}\propto t^{-2}$ in this time interval (\citealt{Margutti19}, their Fig. 9). However, by analogy with AT\,2018cow, we find that optical spectroscopic features emerge in AT\,2024wpp when $L_{\rm{UVOIR}}\approx L_{\rm{X}}$.

The total radiated energy by each emission component is listed in Table  \ref{tab:E_rad}. From this table we note that during the first $\sim 45$\,d, \AT\, radiated $>10^{51}\,\rm{erg}$, a value that is only matched by the most luminous and long-lasting stellar explosions such as superluminous SNe (SLSNe; \citealt{quimby_hydrogen-poor_2011}), and that rules out ordinary SNe for which the \emph{kinetic} energy is $\sim 10^{51}\,\rm{erg}$. As a comparison, AT\,2018cow radiated $\sim 10^{50}\,\rm{erg}$ during the first 60 days of evolution (\citealt{Margutti19}, their Table 1).

\begin{deluxetable}{cc}[h!]
\tablecaption{Energy radiated by \AT\, in the time interval $\delta t=$ 2 $-$ 45\,d \label{tab:E_rad}}
\tablewidth{0pt}
\tablehead{
\colhead{Component} & 
\colhead{Energy Radiated (erg)}
}
\decimalcolnumbers
\startdata
{Swift}-only Blackbody\tablenotemark{a} & $1.11^{+0.03}_{-0.03} \times 10^{51}$\\
Soft X-rays\tablenotemark{b} & $2.8^{+0.1}_{-0.1} \times 10^{49}$\\
$E_{\rm{engine}}\tablenotemark{c}$ & $1.15^{+0.03}_{-0.03} \times 10^{51}$\\
\enddata
\tablenotetext{a}{Using blackbody parameters derived from only the Swift photometry (see Fig.~\ref{fig:BB_fits}).}
\tablenotetext{b}{0.3--10 keV}
\tablenotetext{c}{Derived from $L_{\rm{engine}}$ in Fig. \ref{fig:Lengine}.}
\end{deluxetable}

\section{A persistent, mostly featureless, optical-to-NIR thermal continuum }\label{Sec:featureless}

A distinct observational trait of LFBOTs is the combination of an almost completely featureless optical-to-NIR spectrum with a thermal continuum over a long timescale of weeks after first light. The prominently thermal continuum indicates an optically thick environment, and its persistence with time indicates that this environment is maintained over timescales of weeks. This combination can be obtained with either (i) a large ejecta mass or (ii) slowly expanding ejecta as we show below. For ejecta with mass $M_{\rm{ej}}$, maximum velocity $v_{\rm{max}}$, opacity $\kappa$, in homologous expansion, the optical depth is
\begin{equation}
\tau \approx 160\,  \Big( \frac{M_{\rm{ej}}}{2\,\rm{M_{\sun}}} \Big) \Big(\frac{\kappa}{0.1\,\rm{cm^2/g}} \Big) \Big( \frac{v_{\rm{max}}}{0.3c}\Big)^{-2} \Big( \frac{t}{\rm{day}}\Big)^{-2},
\end{equation}
where we have assumed a constant density profile in radius, and a minimum ejecta velocity $\ll v_{\rm{max}}$. The optically thick condition ($\tau>1$) up to $\delta t\approx 45$\,d either requires fast-moving heavy ejecta with $M_{\rm{ej}}\ge 26\,\rm{M_{\sun}}$ and $v_{\rm{max}} \approx 0.3c$, or light ejecta with $M_{\rm{ej}} \approx 2\,\rm{M_{\sun}}$ and $v_{\rm{max}}\le 20{,}000\,\rm{km\,s^{-1}}$.  A large ejecta mass would violate the constraints from the rapid rise time of AT\,2024wpp, which indicates $M_{\rm{ej}}\lesssim 2\,\rm{M_{\sun}}$,\footnote{We note that this is under the assumption that the rise time tracks the diffusion timescale from a centrally located energy source (as opposed to shock-heated material instead).} while the low maximum expansion velocities are inconsistent with the inferred $R_{\rm{phot}}/t \ge0.2$c (Fig. \ref{fig:BB_fits}).\footnote{In principle, the presence of \emph{pre-existing} material in the transient environment can alleviate this problem, but it would not naturally produce Doppler broadened spectral features with $v\approx$ a fraction of $c$ in the early-time ($\delta t\lesssim10$\,d) spectra.} Another way to put this is that ejecta mass with $\lesssim\,1\,\rm{M_{\sun}}$ expanding at  $v_{\rm{max}}>20{,}000\,\rm{km\,s^{-1}}$ would be optically thin by 45 days, producing optical spectra rich with well-defined spectral features (as in SNe) and thus violating our observations of AT\,2024wpp.  From another perspective, if $\tau (t_{\rm{pk}})\approx v/c\approx (3-5)$ and $\tau(t)\propto t^{-2}$, then we would expect an optically thin spectrum after 1--2 weeks. A solution to this problem is the \emph{continuous} deposition of ejecta mass similar to a wind (as opposed to one episode of ejection, like in an SN), which will keep the optical depth large even if matter is expanding fast and diluting, and/or multiple outflow components with different velocities dominating the detected emission at different epochs.

At the same time, the persistently featureless spectra can be the (combined) result of three main physical scenarios. (i) A steep ejecta density profile above the photosphere, which implies a very small line-forming region, such that the line flux would be negligible and very hard to detect  against the bright, thermal continuum. (ii) Extremely large expansion velocities ($\gtrsim (0.1-0.3)\,c$) that lead to extreme Doppler broadening and line smearing. (iii) Extreme ionization of the ejecta such that recombination is prevented. Option (i) is the main factor leading to the blue and featureless spectra of Type IIP SNe at very early times. However,  in the absence of a central energy source, scenario (i) would lead to the development of strong spectral lines as the ejecta expand and the photosphere recedes in mass coordinates (as observed in SNe), which are not observed in LFBOTs. Interestingly, high ionization has been invoked in the context of TDEs as a way to depress line formation (e.g.,~\citealt{Guillochon2014ApJ, Roth16} and references therein) and, given the observational similarities between TDEs and LFBOTs, might play a role in LBOTs as well.

In the following we thus consider a framework with continuous energy deposition, a luminous central energy source that overionizes the ejecta, combined with extreme Doppler broadening and multiple outflow components as key physical ingredients to explain the observed phenomenology.  A similar model was proposed for AT\,2018cow \citep{Margutti19}, and  we speculate on the astrophysical implications in \S\ref{Sec:Discussion}. The viability of this  model will be quantitatively explored in detail by Aspegren et al., (in prep.) with non-local thermodynamic equilibrium (NLTE) numerical simulations with \texttt{Sedona} \citep{Kasen2006ApJ...651..366K}. 
In this framework, by $t_{\rm{pk}}$, the ``engine'' has deposited $\le 2\,\rm{M_{\sun}}$ of ejecta and the optical/UV emission is 
a combination of reprocessing of X-rays from the central source and thermalization of the  kinetic energy of the outflow (e.g., \citealt{metzger2022aj, tsuna2025}).

\section{The emergence and properties of optical spectral features}\label{Sec:SpectralFeatures}

\begin{figure*}[ht!]
   \centering 
  \includegraphics[width=1.6\columnwidth]{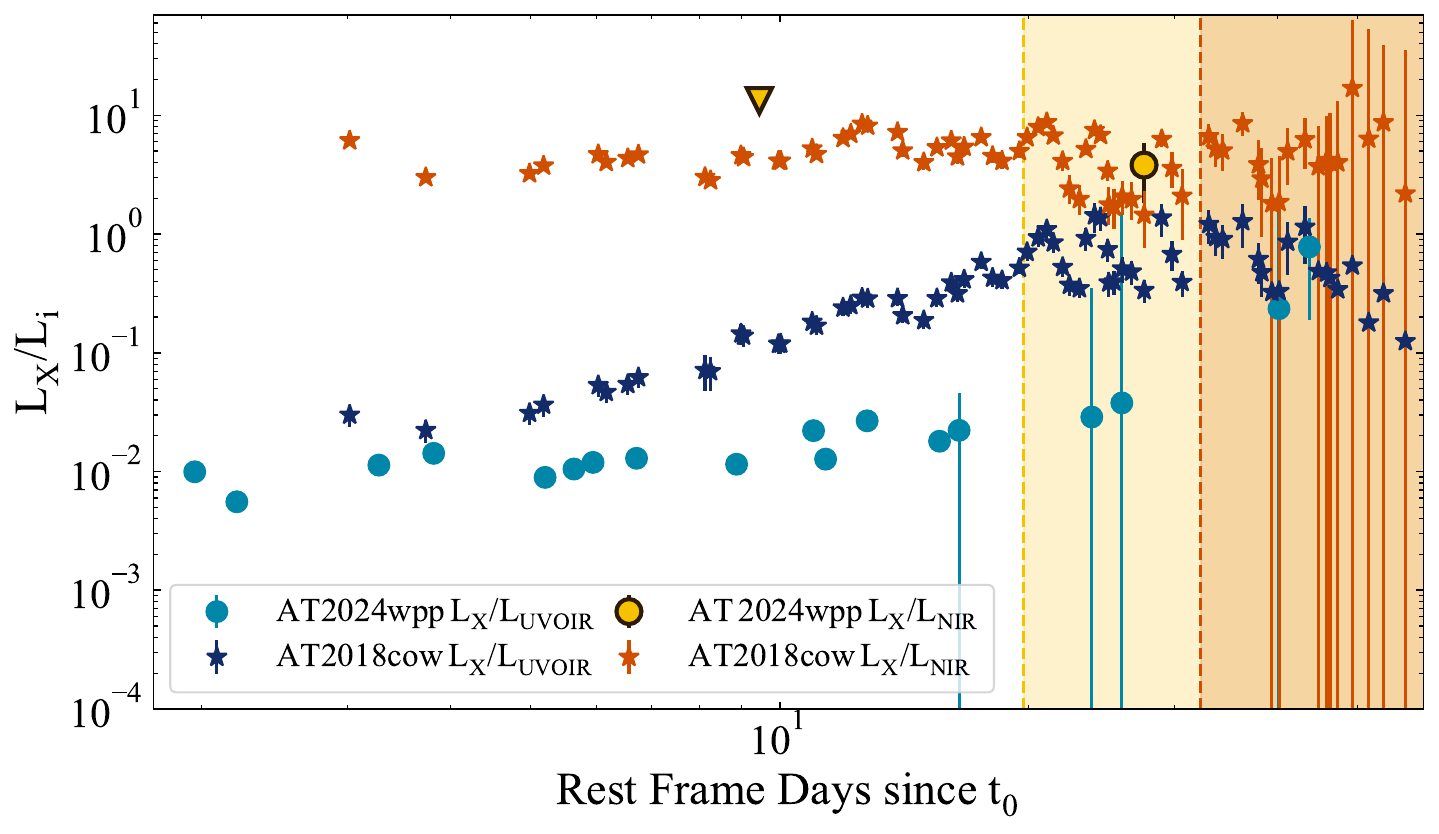}
  \caption{Ratio of the soft (0.3--10 keV) X-ray luminosity ($L_{\rm{X}}$) to $L_{\rm{UVOIR}}$ (shades of orange) and $L_{\rm{NIR}}$ (shades of blue; see \S\ref{Sec:NIRexcess}) for \AT\, (circles) and AT\,2018cow (stars). $L_{\rm{UVOIR}}$ is calculated from the blackbody fits (see Fig.~\ref{fig:BB_fits}). The shaded yellow (orange) area marks the time of emergence of clear spectral features with widths of a few 1000 km s$^{-1}$ in AT\,2018cow (AT\,2024wpp). Interestingly, for both events this happens for $L_{\rm{X}}\approx L_{\rm{UVOIR}}$.  This plot also shows that $L_{\rm{X}}/L_{\rm{NIR}}\approx$ constant with time for AT\,2018cow, and that the parameters of AT\,2024wpp are consistent with the same value. AT\,2018cow data from \cite{Margutti19} and \AT\, $L_{\rm{X}}$ data from Paper II.}
  \label{fig:Lx/Lbol}
\end{figure*}

\subsection{Delayed appearance of spectral features}
\label{SubSec:delayedlines}
Similar to AT\,2018cow (\citealt{Margutti19, metzger2022aj, Piro_Lu_2020, Chen_Shen_2024, Calderon2021MNRAS.507.1092C}), we hypothesize that around the optical peak brightness, the UV-optical emission is dominated by partial reprocessing of the highly variable, inner X-ray source by fast polar outflows (i.e., the \emph{external} shock interaction is subdominant, as supported by the coupled evolution, similar luminosities at late times, and highly variable nonthermal X-ray emission). In this scenario, the temporal evolution of $L_{\rm{X}}/L_{\rm{UVOIR}}$ in Fig. \ref{fig:Lx/Lbol} directly depends on the reprocessing efficiency of the outflow, and hence on its density, temperature, and ionization state (with lower density, higher ionization material having a lower reprocessing efficiency). Figure \ref{fig:Lx/Lbol} shows that $L_{\rm{X}}/L_{\rm{UVOIR}}$ increases from $\sim 0.01$ at optical peak, to $\lesssim 1$ at the time of emergence of clear spectral features with width of a few $1000\,\rm{km\,s^{-1}}$ at $\delta t\approx 30\,$d (Fig. \ref{fig:Line_params}). Interestingly, a similar pattern is followed by AT\,2018cow; however, AT\,2018cow has consistently larger $L_{\rm{X}}/L_{\rm{UVOIR}}$ values and reaches $L_{\rm{X}}/L_{\rm{UVOIR}}\approx 1$ at an earlier stage (with related earlier appearance of the spectral features). This phenomenology is consistent with the observed longer rise time and larger mass of the polar outflow that we inferred for \AT\, (Section \ref{Sec:featureless}) compared to AT\,2018cow. 

\begin{figure}[ht!]
  \centering
  \includegraphics[width=\columnwidth]{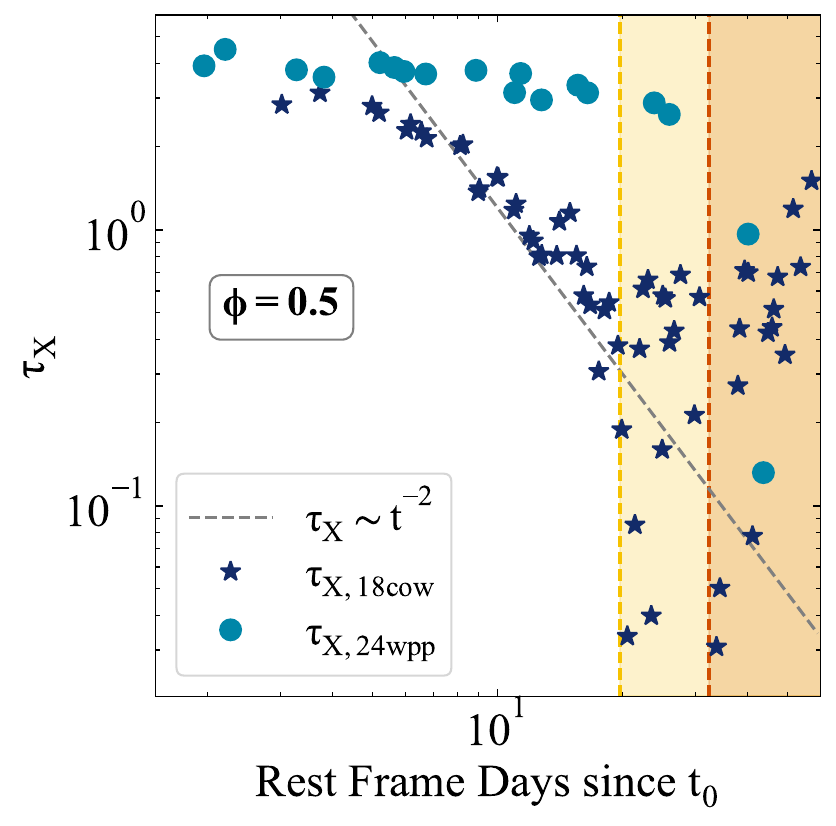}
  \caption{Assuming that the observed $L_{\rm{UVOIR}}$ is purely due to radiative processing of X-rays \citep{metzger2022aj} occurring within (an arbitrary) 50\% of the solid angle (parameterized by $\phi$; in the other half, we assume all X-rays are reprocessed), we calculate the associated $\tau_X$ for \AT\, (circles) and AT\,2018cow (stars) over time and compare to the time of emergence of spectral features (vertical dashed lines; same as Fig. \ref{fig:Lx/Lbol}) for each object. The gray dashed line represents the canonical scaling ($\tau_X\propto t^{-2}$) for radiation escaping a medium expanding at constant velocity. We find that \AT\, maintains a higher $\tau_X$ for longer compared to AT\,2018cow. This is consistent with the features of \AT\  emerging at a later epoch and with the larger inferred $M_{\rm{ej}}$ (\S\ref{Sec:properties}).}
  \label{fig:tau_x}
\end{figure}

\begin{figure}[ht!]
  \centering
  \includegraphics[width=\columnwidth]{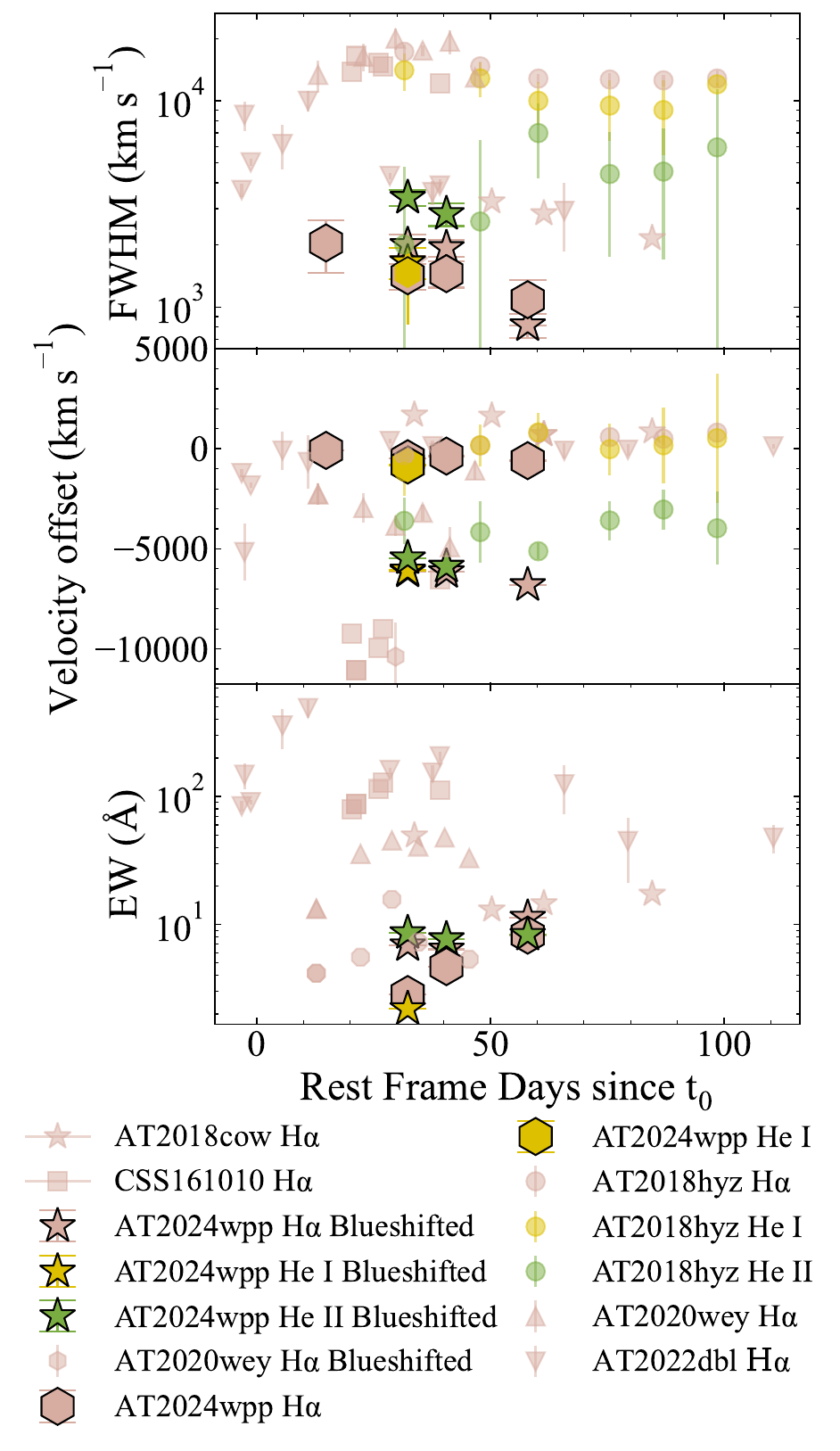}
  \caption{Evolution of the FWHM (top panel), velocity offset (middle panel; i.e., blue or redshift of line peak from the rest wavelength), and line equivalent width (EW; bottom panel) for the H$\alpha$ (pink), He I $\lambda$5876 (gold), and He II $\lambda$4686 (green) emission lines and associated blueshifted components of \AT\, (outlined stars). \AT\, exhibits unique spectral feature evolution compared to LFBOTs AT\,2018cow (non-outlined stars) and CSS161010 (squares), as well as to TDEs AT\,2020wey (upward oriented triangles denote the spectral feature near the rest wavelength and hexagons denote blueshifted components attributed to H$\alpha$ by \citealt{Charalampopoulos2023A&A}), AT\,2018hyz (circles; \citealt{Short2020MNRAS.498.4119S}), and AT\,2022dbl (downward oriented triangles; Guo et al., in prep.). Especially note the slight blueshift in time of the spectral features of \AT. }
  \label{fig:Line_params}
\end{figure}

An increasing ratio $L_{\rm{X}}/L_{\rm{UVOIR}}$ with time is expected as  a result of (i) the expansion of the polar outflow, which leads to lower densities, and/or (ii) increasing ionization of the ejecta \citep{Metzger2014MNRAS.437..703M, tsuna2025}. Both factors are likely at play and lead to a decrease of the effective opacity with time (as shown in Fig.~\ref{fig:tau_x}), thus allowing inner regions to be revealed, while at the same time reducing the effects of line smearing due to photon scattering. The latter effect creates line profiles with broad scattering wings and might make the emission lines effectively undetectable against the continuum at times of larger optical depths, as proposed by \cite{tsuna2025}.  The blackbody radius evolution paints a similar picture.
At early times ($\delta t<6$\,d) we find evidence for an expanding blackbody radius\footnote{We note that in similarity to Type II SNe (e.g.,~\citealt{Rabinak2011ApJ...728...63R}), the blackbody-inferred radius is a better proxy for the photon-thermalization radius instead of the photospheric radius where $\tau_{\rm es}\approx1$.}  (Fig.~\ref{fig:BB_fits}), indicating a brief period of time during which the polar outflow can carry out the photosphere with an inferred velocity of 0.2--0.3$c$. This brief phase is absent in the observations of AT\,2018cow, again consistent with the smaller mass of the polar outflow in this event.
At $\delta t\gtrsim 6$\,d, the inferred radius of the blackbody that best fits the UV-optical emission monotonically decreases with time. The recession of the optical photosphere allows slower moving ejecta components to be revealed and line emission to emerge. In line with this argument, at the time of emergence of spectral features, the H+He line-forming region of \AT{} is roughly at a radius $R_{\rm{line}} \approx  6500\,\rm{km\,s}^{-1}\times \delta t \approx 2\times 10^{15}$\,cm $> R_{\rm BB,\,30\,d}$, under the assumption that the slower moving ejecta was launched at $t_0$.  
We conclude that the  line emission in \AT{} is consistent with originating from an inner region of the ejecta with lower expansion velocities ($\sim 6500\,\rm{km\,s^{-1}}$ vs. $0.2-0.3c$) that is revealed only at later times because  of optical-depth-related effects (e.g., Fig. \ref{fig:tau_x}).

Interestingly, we do not observe a continuum of outflow velocities; rather, in addition to the mildly relativistic outflows with $\sim 0.2-0.3c$, our spectra indicate only two components centered at $6500\,\rm{km\,s^{-1}}$ and $0\,\rm{km\,s^{-1}}$ with similar values of the full width at half-maximum intensity (FWHM) $\approx \rm{few}\,1000\,\rm{km\,s}^{-1}$. This peculiar dual-component profile where the FWHM is less than the blueshift of the centroid is observed for both the H and He features.  Thus, the line-forming regions of both elements (which are likely distinct; see, e.g.,~\citealt{Roth16}) share similar physical conditions and kinematics.  Line formation in an outflowing  medium that is dominated by electron scattering has been demonstrated to lead to blueshifted line profiles (e.g., see models for TDEs such as \citealt{Roth2018ApJ...855...54R}). However, for a homologously expanding outflow, this model leads to a blueshifted component with a profile width that is commensurate with the displacement of the line centroid together with a prominent red wing (e.g., Fig. 6 of \citealt{Roth2018ApJ...855...54R}), which contrasts with our observations of \AT. We thus consider it likely that the line profiles of Fig. \ref{fig:optical_lines}--\ref{fig:optical_halpha} indicate a deviation from spherical symmetry of the emitting region.\footnote{There are known non-purely-kinematic effects that can lead to blueshifted line profiles even in bulk-receding ejecta \citep{vanBaal2023MNRAS.523..954V}. 
However, it is not clear if these physical conditions apply here, and we leave to future work the detailed exploration of this aspect.}
This is not necessarily inconsistent with the low polarization of \AT{} measured between 6--14 d \citep{Pursiainen2025MNRAS} as the weak spectral features indicating asphericity do not appear until at least $\sim20\,$d. Thus, \AT{} may be more spherical at early times before the photosphere recedes and inner ejecta structure is revealed.
Similar to AT\,2018cow, these observations are consistent with a model where 
the H+He emission originates from lower-velocity equatorial material, with polar outflows carrying the mildly relativistic ejecta (\citealt{Margutti19}, their Fig.~12; Paper II, Fig.~14). These conditions are naturally realized in super-Eddington accretion disks (e.g.,~\citealt{Sadowski15,Sadowski&Narayan2016MNRAS.456.3929S}). Interestingly, radiation-hydrodynamic simulations of super-Eddington accretion flows around BHs by \cite{Yoshioka24} found evidence for two components of outflows, with a faster, lighter component having velocity $\gtrsim 0.1c$ ejected along the polar direction (i.e., within $\sim 10\degree$), and slower, denser outflows having typical velocities of a few $1000\,\rm{km\,s^{-1}}$ at larger angles (their Fig. 4). The presence and properties of these two outflow components are consistent with the observations of \AT\, and LFBOTs that have detailed spectroscopic sequences.

In Fig.~\ref{fig:Line_params}, we compare the evolution of the FWHM, velocity offset from line center, and equivalent width (EW) of \AT's spectral features to other transients. To date, CSS161010 is the only other FBOT to show blueshifted spectral features;  these features are up to a factor of 10 broader than those observed in \AT\, and are blueshifted by $\sim 10{,}000\,\rm{km\,s^{-1}}$, other than the final epoch at $40\,$d which is blueshifted by $\sim 6500\,\rm{km\,s^{-1}}$, similar to \AT's features. Neither \AT's nor CSS161010's line profiles show the red wing expected for an electron-scattering dominated outflowing medium (\citealt{Roth2018ApJ...855...54R}). 
A few TDEs show blueshifted spectral features: AT\,2020wey possibly has blueshifted secondary H$\alpha$ peaks in several optical spectra \citep{Charalampopoulos2023A&A}, though the spectra have low S/N; PTF09ge \citep{Arcavi2014ApJ...793...38A}, 
ASASSN-15oi \citep{Holoien2016MNRAS.463.3813H}, 
and SDSS J0748 \citep{Wang2011ApJ...740...85W} 
all exhibit blueshifted He II $\lambda4686$ in at least one epoch; 
and ASASSN-14ae \citep{Holoien2014MNRAS.445.3263H} 
and AT\,2022dbl (Guo et al., in prep.) both show a blueshift of H$\alpha$ in their earliest spectrum. 
Many of these TDE features can be at least partly explained by electron scattering \citep{Roth2018ApJ...855...54R}, potentially alongside Bowen fluorescence feature blends for the He II profiles (see  \citealt{Gezari2015ApJ...815L...5G, Brown2018MNRAS.473.1130B}). We also note that TDE features tend to have much larger EWs than those measured for \AT.

\subsection{Line Luminosity}
\label{SubSec:lineluminosity}

Another peculiarity of the spectral features of \AT\, is their low line luminosity and EW. Peaking at $L_{\rm{H\alpha}}\approx 10^{39}\,\rm{erg\,s}^{-1}$, the observed line luminosity  is $\sim0.01\%$ of the bolometric luminosity at the same time. We follow the reasoning by \citet[][their Section 4.2]{tsuna2025} and consider a photoionization origin for the detected lines. For comparison, shock ionization of $1\,\rm{M_{\sun}}$ of material would contribute at most $\sim 4\times 10^{45}\,\rm{erg}$ in H$\alpha$ line assuming 100\% efficiency, and we observed $\gtrsim 10^{45}\,\rm{erg}$. We thus consider shock ionization less likely. The amount of ionized mass ($M_{\rm{ion}}$) in the slow-moving ejecta can either be the total mass (density-bounded regime) or lower (ionization-bounded regime; e.g., \citealt{Osterbrock2006agna.book.....O}).

In the ionization-bounded regime, the H$\alpha$ line luminosity is (\citealt{tsuna2025}, their Eq. 50)
\begin{equation}
L_{\rm{H\alpha}}\approx \dot N_{\rm{ion}}\epsilon_{\rm{H\alpha}} \Big (  \frac{\alpha_{B}^{\rm{H\alpha}}}{\alpha_{B}}  \Big )\, ,
\label{Eq:LHalpha}
\end{equation}
where $\dot N_{\rm{ion}}\approx \Phi {L_{\rm{ion}}}/{\epsilon_{\rm{ion}}}$ is the photoionization rate; $L_{\rm{ion}}$ is the ionizing luminosity; $\Phi $ is the fraction of ionizing luminosity intercepted by the slow ejecta; $\epsilon_{\rm{ion}}$ is the energy ``cost'' for each hydrogen ionization, which depends on the details of photoionization and recombination of each species in the ejecta (here we use $\epsilon_{\rm{ion}}\approx 30\,\rm{eV}$ following \citealt{tsuna2025}); $\alpha_B$ is the recombination coefficient and $\alpha_B^{\rm{H\alpha}}$ is the H$\alpha$ recombination coefficient; and $\epsilon_{\rm{H_{\alpha}}}$ is the H$\alpha$ photon energy. For a typical recombination branching fraction of $\alpha_B^{\rm{H\alpha}}/\alpha_B\approx 1/3$, Eq. \ref{Eq:LHalpha} leads to the following H$\alpha$ production efficiency
\begin{equation}   
\frac{L_{\rm{H\alpha}}}{L_{\rm{ion}}}\approx 10^{-2} \Big ( \frac{\Phi}{0.5}\Big ) \Big (\frac{\epsilon_{\rm{ion}}}{30\,\rm{eV}}\Big )^{-1}\, .
\end{equation}
The significantly lower ratio observed, ${L_{\rm{H\alpha}}}/{L_{\rm{engine}}}\gtrsim 10^{-4}$, may be due to a significantly lower $\Phi$ (i.e., geometric effects), a small fraction of $L_{\rm engine}$ being ionizing photons, a H-depleted slow outflow (e.g., a tidally disrupted WR star), or that this is instead in the density-bounded regime.

The maximum ionized mass is given by Eq.~49 from \cite{tsuna2025},
\begin{equation}
\begin{split}
    M_{\rm{ion, max}} \approx \, & m_p\sqrt{\frac{4\pi R_{\rm line}^3 \Phi L_{\rm rad}/\epsilon_{\rm ion}}{3\alpha_B}} \approx 0.2\,\mathrm{M_{\odot}} \Big (\frac{\Phi}{0.5} \Big )^{1/2}\\ \left(L_{\rm rad}\over 10^{43}\mathrm{erg/s}\right)^{1/2}
    & \Big ( \frac{\epsilon_{\rm{ion}}}{30\,\rm{eV}} \Big )^{-1/2}
    \Big ( \frac{R_{\rm line}}{2\times 10^{15}\,\rm{cm}} \Big )^{3/2} \Big ( \frac{T}{10^{5}\,\rm{K}}\Big )^{0.35}\, ,
\end{split}
\label{Eq:Mion}
\end{equation}
where we used a fiducial value of $R_{\rm line} \approx 6000\mathrm{\,km/s} \times \delta t \approx 2\times10^{15}\rm\, cm$ for the radius relevant to the slower ejecta of \AT.

If the actual mass of the ionized H$\alpha$ emitting gas, $M_{\rm ion}$, is much lower than $M_{\rm{ion,max}}$, we would be in the density-bounded regime and hence the line production efficiency $L_{\rm{H\alpha}}/L_{\rm{rad}}$ will be lower than that in the ionization-bounded regime ($\sim1\%$) by a factor of $M_{\rm ion}/M_{\rm ion,max}$. In the super-Eddington disk outflow picture, the majority of the mass is carried by the slowest outflow that originates from the outer disk. For an outer disk radius of $R_{\rm d}$ and compact object mass $M$, we expect the velocity of the slowest outflow to be $v_{\rm min}\sim \sqrt{GM/R_{\rm d}}$. The outflow near $v_{\rm H\alpha}\approx 6000\rm\, km\,s^{-1}$ is launched from smaller radii near $(v_{\rm min}/v_{\rm H\alpha})^2 R_{\rm d}$, and hence the outflow mass near velocity $v_{\rm H\alpha}$ is roughly given by $M_{v_{\rm H\alpha}} \sim (v_{\rm min}/v_{\rm H\alpha})^{2p} M_{\rm d}$, where $M_{\rm d}$ is the total disk mass and we have adopted a radial power-law scaling for the accretion rate in a super-Eddington disk $\dot{M}(r)\propto r^p$. For $p=0.5$ \citep[as adopted by][]{tsuna2025}, we expect
\begin{equation}
    M_{v_{\rm H\alpha}}\sim 0.03 M_\odot {(M_{\rm d}/M_\odot)(M/1.4 M_\odot)^{1/2}\over (R_{\rm d}/10R_\odot)^{1/2}},
\end{equation}
where we have adopted fiducial values of $M_{\rm d}\sim M_\odot$ and outer radius $R_{\rm d}\sim 10 R_\odot$ as the disk loses mass and viscously spreads over time (assuming that the line emission comes from the freshly launched slow outflow). We find that the system may indeed be in the density-bounded regime, but this only reduces by line production efficiency by an order of magnitude to about 0.1\% if $\Phi\sim 0.5$. Geometric effects ($\Phi\ll 1$) or a hydrogen-poor outflow may further reduce the line production efficiency to the observed value.

To conclude and summarize, the structure of the spectral features of \AT\, points to an ejecta geometry with clear deviation from spherical symmetry involving multiple outflows including a fast, collimated component as well as a slower component. Outflows from disks accreting at super-Eddington rates are a plausible formation scenario of the observed dual line-profile structure (see, e.g., \citealt{Yoshioka24}).

\section{A NIR excess of emission}\label{Sec:NIRexcess}

\begin{figure*}[ht!]
  \begin{subfigure}
    \centering
    \includegraphics[width=2\columnwidth]{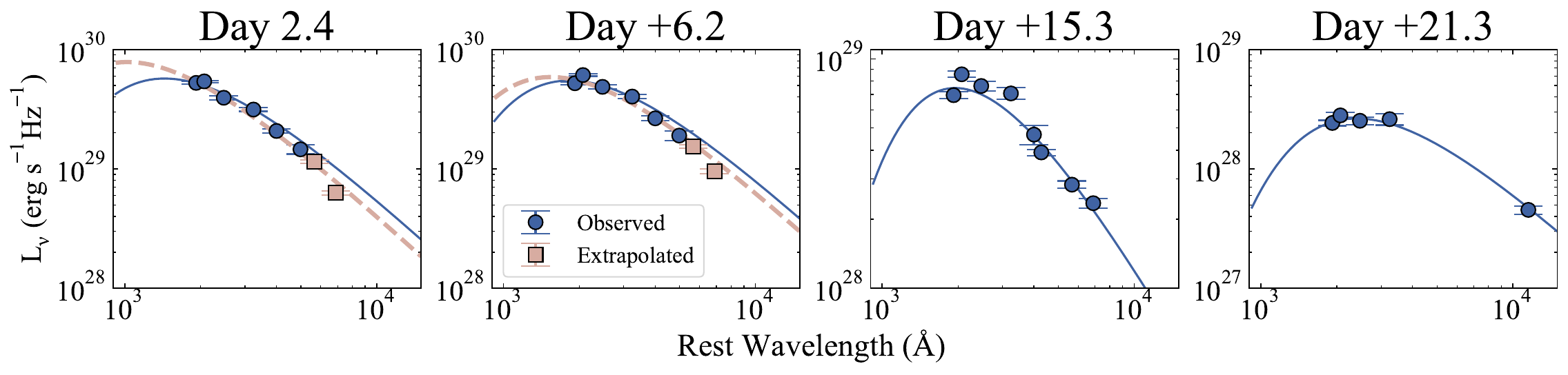}
    \end{subfigure}
    \begin{center}
    \begin{subfigure}
    \centering
    \includegraphics[width=1.5\columnwidth]{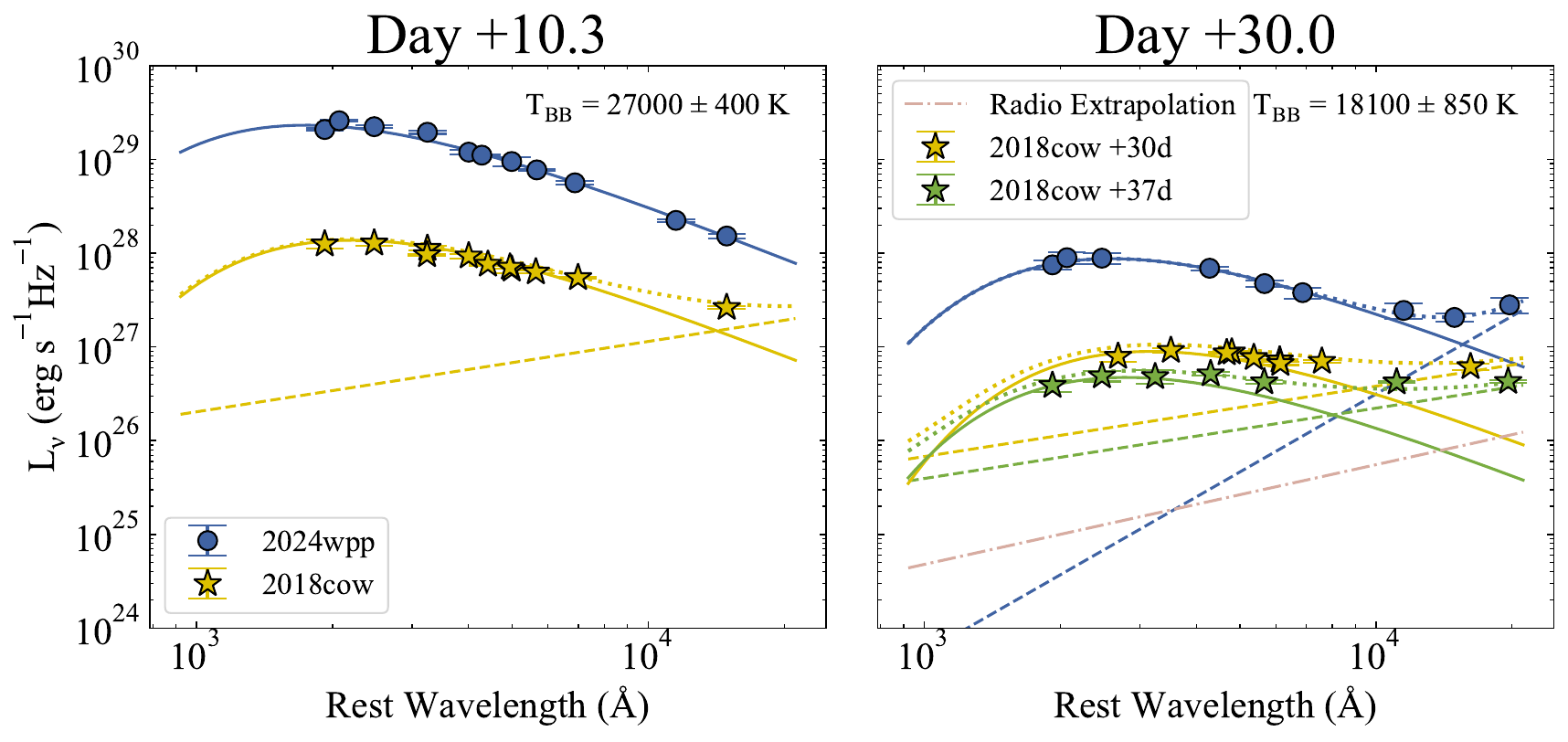}
    \end{subfigure}  
    \end{center}

  \caption{\textit{Top Panel:} Blackbody fits to the observed photometry (blue points) at various epochs (observed frame days). Pink points are extrapolated \textit{r}- and \textit{i}-band photometry that assume the {$w1$} -- \textit{r} and {$w1$} -- \textit{i} colors shown in Fig.~\ref{fig:color}.  
  \textit{Bottom Panel}: \AT\, SEDs on the (observed frame) days of its $J$- and $H$-band photometry compared to coeval SEDs of AT\,2018cow. We find no evidence of emission in excess of a blackbody fit (blue curves) until day +30, at which time a blackbody plus power-law model (blue dashed curve) is required to fit the SED. The best-fit power law (blue dashed line; F$_\nu \propto  \nu^{-\alpha}$) has $\alpha = 2.7 \pm 0.5$. \AT\, is not fit well by the power law ($\alpha=0.75$ yellow and green dashed lines) that \citet{perley2019mnras} and \citet{chen2023a} used to fit the NIR excess of AT\,2018cow. Additionally, the extrapolation of the coeval radio SED (pink curve; from Paper II) cannot explain the observed NIR excess. 
  }
  \label{fig:SED}
\end{figure*}

The appearance of a NIR excess of emission at early times is a hallmark observational feature of LFBOTs\footnote{However, we note that similar NIR excesses have also been observed in SNe that are shrouded in very dense surrounding media (e.g., SN\,2009ip \citealt{margutti2014a}, their Fig. 27; \citealt{Smith2013MNRAS}, their Fig. 2).}. We first summarize the key observational facts, and then consider two interpretations: reprocessed radiation from a pre-existing dust shell (i.e., a dust echo in \S\ref{SubSec:NIRDustEcho}) and frequency-dependent opacity effects due to free-free processes in  the transient's outflows or pre-existing material (\S\ref{SubSec:NIRfreefree}).

The NIR observational results are as follows. 
\begin{itemize}
    \item We find evidence for a deviation from the blackbody that best fits the UV-optical SED at NIR wavelengths. At $\delta t =30.0$\,d, the local power-law slope measured from our photometry  in the NIR is $F_{\nu,\rm{NIR}}\propto \nu^{-\alpha}$ with $\alpha\approx 0.3$. Following \citet{perley2019mnras} and \citet{chen2023a}, we fit the overall spectrum with a phenomenological model consisting of a blackbody + power law ($F_{\rm\nu,BB}+F_{\rm\nu,PL}$), finding  $F_{\nu,PL}\propto \nu^{-3}$ (Fig.~\ref{fig:SED}). 
    
    \item The NIR luminosity of \AT\, at (rest frame) +27.6 d determined from $F_{\nu,PL}$ (integrated in the range $2300$\,\AA\,-- $2.4\,\mu$m) is $L_{\rm{NIR}}= (1.9\pm0.3) \times 10^{41}$ $\rm{erg}\,\rm{s}^{-1}$, which is comparable to the power-law luminosity of AT\,2018cow at (rest frame) +29.8 d ($L_{\rm{NIR}}^{\rm{18cow}}= 3.0 \pm 0.6 \times 10^{41}$ erg\,s$^{-1}$ over the same wavelength region, using the power-law parameters from \citealt{chen2023a}). 
\item The NIR ``excess'' is clearly detected at $\delta t =30.0$\,d, but it was not detected in our previous epoch with NIR sampling at day +10.3, nor in the final \textit{z}-band epoch (day +17.9). Our Keck NIR spectroscopy at 24.2\,d indicates $F_{\nu,\rm{NIR}} \propto \nu^{-\alpha}$ with $\alpha>0$ (especially at wavelengths $>1.9\,\mu\rm{m}$; see Fig.~\ref{fig:NIR_spec}) consistent with the day +30 findings, suggesting that the NIR excess was already emerging at $\delta t =20$\,d (this aligns with the 20\,d NIR excess reported by \citealt{Pursiainen2025MNRAS}). 
At $\delta t =10.3$\,d we do not find significant evidence for a departure from a blackbody spectrum in the NIR, and we estimate a 3$\sigma$ limit on the (undetected) NIR excess of $L_{\rm{NIR}}<5.6\times 10^{41}\rm{erg\,s^{-1}}$. 
We note that the later detection of a NIR excess in \AT\, (20--30\,d) compared with AT\,2018cow, which showed a NIR excess within a few days, can be related to the significantly larger UV-optical luminosity of \AT.
\item The ratio of the soft X-ray luminosity (0.3--10 keV) to the NIR power-law luminosity ($L_{\rm{x}}/L_{\rm{NIR}}$) for \AT\, is consistent with that of AT\,2018cow, remaining constant with time at $\sim 0.5$ (Fig.~\ref{fig:Lx/Lbol}). 
\item As in AT\,2018cow \citep{Margutti19}, the extrapolation of the radio spectrum of AT\,2024wpp (from Paper~II) to the NIR range significantly underpredicts the NIR flux (Fig.~\ref{fig:SED}), even in the absence of any spectral break in between the two wavelength regimes, implying that the NIR excess is not directly connected with the  nonthermal radio synchrotron spectrum. 
\end{itemize}

\subsection{Dust Echo} \label{SubSec:NIRDustEcho}

Following \citet{Metzger&Perley2023} and \citet{tuna2025}, we investigate whether the NIR excess can be explained by reprocessing of early UV emission by a pre-existing, opaque, dusty medium. \citet{Metzger&Perley2023} postulate that LFBOT progenitors may be surrounded by a dense and cool CSM at large radii ($\gtrsim 10^{16}$ cm) where dust can form before the LFBOT. This ``dusty outer shell'' would 
initially be opaque to UV photons until reaching sublimation temperatures ($\sim2000$\,K).
In this model the absorbed energy is reradiated as a NIR ``echo'' on a timescale mostly set by 
geometric time delays. The persistent NIR excess of AT\,2018cow,  observed during $\delta t = 3-44$\,d,  was interpreted by \citet{Metzger&Perley2023} as a dust echo originating from a medium with density $n(r)=n_0(r/r_0)^{-3}$ and  $n_0\approx 3\times10^7$\,cm$^{-3}$ at $r_0=10^{16}\,\rm{cm}$ (assuming 1\,$\mu$m silicate grains and a dust-to-gas ratio ${X_d}=0.1$). We note that the assumed $n(r)$ scaling was motivated by the findings from the radio modeling. 
This model is appealing as the derived medium density is not too different from that inferred from the radio modeling of AT\,2018cow if large deviations from equipartition are considered \citep{Margutti19}.

A similarly steep and dense CSM density profile is obtained from the radio modeling of \AT\, (Paper\,II, their Fig.~9), motivating us to explore the dust-echo model for \AT. Under the same model assumptions, the maximum duration of a NIR excess of emission is (\citealt{Metzger&Perley2023}, their Eq. 22) 
\begin{multline}
    \Delta t_{\rm IR,max} \approx  31.6 \rm{~d~} \times \\\left(\frac{n_0}{10^7\, \rm{cm^{3}}}\right)^{1/2} \left(\frac{r}{10^{16}\, \rm{cm^{3}}}\right)^{-3} \left( \frac{X_d}{0.1}\right) \left(\frac{a}{1\,\mu {\rm m}}\right)^{-1/2}\, .
\label{t_IR_max}
\end{multline}
\noindent
The NIR excess in \AT\, lasts $\ge 30$ d and thus implies $n_0\gtrsim 10^{7}\,\rm{cm^{-3}}$, similar to AT\,2018cow. As a comparison, the CSM density inferred from the radio modeling of \AT\, in equipartition is $n\approx1\times 10^5\,\rm{cm}^{-3}$ at $10^{16}\,\rm{cm}$ (see Paper II).  Similar to above, this density can only be comparable with the density of the NIR-emitting material derived above \emph{if} corrections from a large deviation from equipartition are applied.

The predicted NIR luminosity (Eq. 23 of \citealt{Metzger&Perley2023}, here renormalized using the peak bolometric luminosity $L_{\rm{pk}}$ and rise time $t_{\rm{pk}}$ of \AT) is
\begin{multline}
\label{eqn:LNIR_dust}
    L_{IR} \approx 3.2 \times 10^{40}\, \rm{erg~s^{-1}} \left(\frac{n_0}{10^7\, \rm{cm^{-3}}} \right ) \left(\frac{r}{10^{16}\, \rm{cm^{3}}}\right)^{-3} \\  \left( \frac{X_d}{0.1} \right ) \left(\frac{a}{1\,\mu {\rm m}}\right)^{1/2}
    \left(\frac{L_{\rm{pk}}}{2\times10^{45} \rm{~erg~s^{-1}}}\right)^{-1/2} \left( \frac{t_{\rm{pk}}}{4 \rm{~d}}\right )\, .
\end{multline} 

The observed $L_{\rm{NIR}}=(1.9\pm0.3) \times 10^{41}$ erg\,s$^{-1}$ thus implies $n_0\approx 4\times 10^{7}\,\rm{cm^{-3}}$, again consistent with the radio-inferred CSM density for large deviations from equipartition. In this scenario, while the NIR and the radio do \emph{not} belong to the same component of emission, they can nevertheless originate from the same medium.\footnote{We note that Equations \ref{t_IR_max} and \ref{eqn:LNIR_dust} only apply to optically \emph{thick} dust and that radiation will continue to get reprocessed to IR wavelengths by optically thin dust beyond the maximum dust sublimation radius \citep{tuna2025, li2025}. This effect likely contributed to the $>30$\,d NIR excess observed in AT\,2018cow which was not well fit by the opaque dust models explored by \cite{tuna2025}.}  

\subsection{Free-Free Opacity Effects in an Extended ``Atmosphere''} \label{SubSec:NIRfreefree}

\citet{Chen_Shen_2024} propose that an LFBOT NIR excesses can be related to free-free opacity effects occurring within an extended medium having a shallow density profile above the \emph{optical} photosphere $r_{\rm{ph}}$. In this region at $r>r_{\rm{ph}}$, the absorptive opacity $\kappa_a$ changes systematically with frequency --- that is, different frequencies of the continuum spectrum will have different effective thermalization radii, producing a deviation from a single-temperature blackbody spectrum. This is analogous to the process that produces NIR/radio excesses in hot stars surrounded by dense winds (e.g., \citealt{Wright1975MNRAS.170...41W,Crowther2007ARA&A..45..177C}), and it has been suggested by \citet{Roth16} to play a major role in shaping the continuum in TDEs.  Here we follow \citet{Roth16}, \citet{lu2020mnras}, \citet{Chen_Shen_2024}, and \citet{Somalwar2025}, and assume that free-free dominates the absorptive opacity ($\kappa_a\approx \kappa_{\rm{ff}}$), and that in the  extended atmosphere $\kappa_{\rm{ff}}\ll \kappa_{\rm{es}}$ (where $\kappa_{\rm{es}}$ is the electron-scattering opacity), such that the effective opacity is $\kappa_{\rm{eff}}\approx (\kappa_{\rm{ff}}\kappa_{\rm{es}})^{1/2}$ (see \citealt{RadiativeProcesses}, Ch.~1). In this treatment the medium is assumed to be effectively optically thick.
We generalize the \citet{lu2020mnras} analytical model for a power-law density profile  of the emitting medium above the optical photosphere $r_{\rm{ph}}$ of the form  $\rho(r)=\rho_0(r_0/r)^{-s}$.
The specific luminosity is $L_\nu\approx 4\pi j_\nu V$, where $j_\nu$ is the emissivity; the emitting volume is 
$V=f r_{\rm th,\nu}^3$ for a frequency-dependent thermalization radius $r_{\rm{th,\nu}}$,
\begin{multline}
    r_{\rm{th},\nu}= (0.018 \kappa_{\rm{es}})^{\frac{1}{3s-2}} \left (\frac{\rho_0^{3/2} r_0^{3s/2} Z} {m_p \mu_e \mu_I^{1/2} (s-1)} T^{-3/4}\nu^{-1}\right)^{\frac{2}{3s-2}}\, ,
\end{multline}
where $T$ is the temperature of the medium, $m_p$ is the proton mass, $Z$ is the atomic number, and $\mu_e$ ($\mu_I$) is the mean molecular weight for electrons (ions). 
The expected spectrum scales as $L_{\nu}\propto \nu^{\frac{4s-6}{3s-2}}$   for $s>1$ (\citealt{Margutti19}, their Eq. 6). We report below the entire $L_{\nu}$ expression for completeness: 
\begin{multline}
\label{eqn:LNIR_freefree}
    L_{\nu} = \frac{8\pi}{\kappa_{\rm es} c^2} k_B (s-1) f \left(\frac{0.018 \kappa_{\rm es}}{m_p^2 (s-1)}\right)^{\frac{s+1}{3s-2}} \\ \mu_e^{\frac{s-4}{3s-2}}\mu_I^{\frac{s+1}{2-3s}} Z^{\frac{2+2s}{3s-2}}\nu^{\frac{4s-6}{3s-2}} T^{\frac{3s-7}{6s-4}} \rho_0^{\frac{5}{3s-2}} r_0^{\frac{5s}{3s-2}}\, .     
\end{multline} 
For $s=2$, we find the well-known wind-like medium scaling $L_{\nu}\propto \nu^{1/2}$.

The  $L_{\nu,\rm{NIR}} \propto \nu^{-0.3}$ observed at $\sim 30$\,d is broadly consistent with an $s\approx 1.3$ medium. For this medium, the observed NIR $L_{\nu}$ requires $\rho_0\approx 7\times 10^{-16}\,\rm{g\,cm^{-3}}$ for an assumed $r_0 = 10^{16}\,\rm{cm}$, $T\approx 20{,}000\,$K, $f=4\pi$ (i.e.,~for spherical symmetry), and H-dominated composition. The derived density  is $\sim 10^4$ times the CSM density inferred from the radio modeling in equipartition.  Thus, in this scenario, the radio and NIR emission components are unlikely to originate from the same CSM and the NIR originates from extended material above the optically emitting medium. 
With a density profile $\propto r^{-1.3}$, the medium is significantly shallower than a wind-like profile $\rho = \dot{M}/4\pi v_{w}r^2 $ that is expected for constant $\dot M /v_w$. For an assumed wind velocity $v_w = 0.2c$ (motivated by the ``expansion velocity'' of the blackbody; see Fig. \ref{fig:BB_fits}), the density above corresponds to an effective mass-loss rate $\dot M_{\rm{eff}}\approx 80\,\rm{M_\odot\,yr^{-1}}$. If maintained over a month timescale, this value implies a total ejecta mass of $\sim7\,\rm{M_\odot}$.
For these parameters we calculate $r_{\rm th,\nu}\approx 2 \times 10^{15}\,\rm{cm}$ at $\nu_{\rm{NIR}}\equiv 10^{14}\,\rm{Hz}$ at 30\,d,
which is a factor of $\sim4$ larger than the blackbody radius ($r_{\rm BB}\approx4\times10^{14}\,\rm{cm}$).
The  mass in the region $r_{\rm BB}<r<10^{16}\,\rm{cm}$ at 30\,d is $\sim 2.5$\,M$_{\odot}$, which is similar to the ejecta mass inferred with Eq. \ref{eqn:ejectamass}. 
We note that it is quantitatively unclear whether very large velocities and the presumably high level of ionization are enough to prevent the detection of prominent lines from this material above the optical photosphere, and we leave the detailed investigation of this aspect to future work. 

\citet{Chen_Shen_2024} find this model can well explain the optical-to-NIR emission in AT\,2018cow using a wind-like medium scaling for $L_\nu$. They infer at early times ($\delta t<10\,$d), $\dot{M}\approx60\,\rm{M_\odot\,yr^{-1}}$, and at late times,  $\dot{M}\approx22\,\rm{M_\odot\,yr^{-1}}$, with $\dot{M}(t)\propto t^{-1.8}$. Over the 15\,d evolution of AT\,2018cow, \citet{Chen_Shen_2024} estimate the outflow total mass to be $\sim5.7$\,M$_\odot$. These parameters are similar to our inferences for AT\,2024wpp. 

\,\,\,\,\,\,\,\,\,\,\,\,\,\,\,\,\,\,\,\,\,\,\,\,\,\,\,\,\,\,\,\,\,\,\,\,\,\,\,\,\,\,\,\,\, \,\,\,\,\,\,\,\,\,\,*\,\,\,\,\,    *  \,\,\,\,  *

We end with two considerations. First, neither the dust-echo nor the free-free opacity effects models provide  
a natural explanation for $L_{\rm X}/L_{\rm NIR}\approx0.5$ for both AT\,2018cow and \AT, nor for the constancy of this ratio throughout the evolution of AT\,2018cow (Fig. \ref{fig:Lx/Lbol}). Observations of a larger sample of LFBOTs that extend to the NIR will clarify if this is a peculiarity of the two events known, or a failure of current models. 
Interestingly, we note that the LFBOT-like transient AT\,2024puz also exhibits a NIR excess that was interpreted in the context of both of these models \citep{Somalwar2025}.

 Second, as inferred for AT\,2018cow by \citet{Margutti19}, for very steep density profiles $s\gg 1$, Eq. \ref{eqn:LNIR_freefree} asymptotically converges to  $L_{\nu}\propto \nu^{4/3}$. This is not dissimilar from the measured slope of the early optical-UV spectrum of AT\,2024wpp at $\delta t\approx$ 6\,d ($L_{\nu}\propto \nu^{1.3}$), implying that the \emph{optical} continuum is formed in a medium with a steep density gradient, as in AT\,2018cow. This is important, since the steepness of the optical continuum-forming medium is a key physical ingredient that we identified in \S\ref{Sec:featureless} to obtain featureless spectra. We note that this last inference is independent of the explanation of the NIR excess. 

\section{Discussion}\label{Sec:Discussion}
\subsection{Comparison of \AT\, with Other Transients}

In the previous sections we have compared \AT{} with ``classical,'' ``cow-like'' LFBOTs. Here we expand our comparisons to more broadly include classes of transients that share some aspects of LFBOT phenomenology (in particular, fast evolving luminous transients and TDEs).

\textbf{Fast Evolving Luminous Transients:} AT\,2024puz \citep{Somalwar2025} is a peculiar transient with similar properties to both LFBOTs and TDEs (e.g.,~persistent blue colors, luminous optical emission $L_{\rm{pk}}\approx 10^{44.8}\,\rm{erg\,s}^{-1}$, featureless spectra, bright and highly variable X-ray emission $L_{\rm{X}}\approx 10^{44.1}\,\rm{erg\,s}^{-1}$, NIR excess emission above a blackbody), but the UV-optical light curve evolves on an intermediate timescale (10\,d rise time, $\sim 20$\,d evolution timescale; slow for LFBOTs, fast for TDEs). \citet{Somalwar2025} interpret this object as a slowly evolving LFBOT. The discovery of it  points to a continuum of phenomena between LFBOTs and TDEs, and suggests that future LFBOT searches may benefit from including phenomena that evolve on slightly longer timescales (\citealt{Somalwar2025}).  

Other fast-rising and fading ($\sim 10\,$d), luminous ($M_{\rm peak}<-20\,$mag) transients such as Dougie \citep{Vinko2015ApJ}, AT\,2022adem \citep{nicholl_at_2023}, and AT\,2020bot \citep{ho2023a,nicholl_at_2023} may also exist within this LFBOT to TDE continuum. Both Dougie and AT\,2022adem have hot, blue spectral continua, but while Dougie remains featureless through $30\,$d, AT\,2022adem shows H+He emission beginning from 3\,d. These features transition to absorption after 14\,d.
AT\,2020bot may be spectrally different from the former two objects with broad, weak lines at peak brightness that are distinct from traditional SN features and are not obviously identifiable.
All of these transients exhibit much faster cooling of their optical-UV emission than typical LFBOTs (especially AT\,2020bot) and all occurred in elliptical galaxies with low recent star formation \citep{nicholl_at_2023}, unlike typical (L)FBOT host galaxies \citep{ho2023a}. Additionally, Dougie and AT\,2022aedm do not show the luminous X-ray/radio emission that is characteristic of LFBOTs (AT\,2020bot was not observed at X-ray or radio wavelengths).  

\textbf{TDEs:} There are clear observational analogies between TDEs and LFBOTs. Both classes of objects are characterized by persistently blue optical light curves \citep{vanVelzen2011ApJ...741...73V,vanvelzen2021a,Arcavi2014ApJ...793...38A}, and typically only show signatures of H and He in their spectra \citep{Hammerstein2023a, Yao2023ApJ...955L...6Y, vanvelzen2021a}. The potentially connected subclasses of featureless and jetted TDEs both have persistent featureless optical spectra \citep{Andreoni2023Natur.613E...6A, Hammerstein2023a, Yao2023ApJ...955L...6Y}. As discussed in \S\ref{Sec:featureless}, featurelessness can be attributed to a combination of a steep density profile limiting the size of the line-formation region, large expansion velocities leading to extreme line broadening, and high ionization of the ejecta.

By analogy with TDE literature (e.g.,~\citealt{Guillochon2014ApJ,Roth16}), we lean toward the latter two explanations for the  featurelessness of \AT\, and LFBOTs. We posit the presence of a reprocessing envelope (as in some models of TDEs) but remain agnostic about the astrophysical origin \citep{LoebUlmer1997ApJ...489..573L, Metzger2022TDE}. TDEs are typically considered nuclear transients involving a supermassive BH  (SMBH)  (\citealt{vanVelzen2019ApJ...872..198V,vanvelzen2021a,Hammerstein2023a,Yao2023ApJ...955L...6Y}; though recently an off-nuclear SMBH TDE was discovered --- see \citealt{Yao2025ApJ}). Most LFBOTs (including \AT) are decidedly off-nuclear transients \citep{chrimes_at2023fhn_2024}, thus a TDE-like interpretation likely requires an IMBH or stellar-mass BH (or NS). Along with LFBOTs, TDEs are the only other transients known to have late-time X-ray/UV plateaus \citep{Mummery2024MNRAS.527.2452M} as observed for AT\,2018cow \citep{migliori2023ae, chen2023a, Chen2023b, inkenhaag2023mnras, Sun2022MNRAS.512L..66S, Sun2023MNRAS.519.3785S}. For both classes of transients, this late-time emission is attributed to an accretion disk, but as LFBOTs involve much less massive compact objects, their accretion rate is likely (highly) super-Eddington, which is not always true in TDEs. Thus, LFBOTs may allow us to probe objects with much higher accretion rates than typical TDEs.

\subsection{LFBOT Models}
\label{SubSec:LFBOTmodels}
\AT\, radiated an extreme $\sim10^{51}\,\rm{erg}$ over the first $\sim 45$\,d, surpassing AT\,2018cow in the same time period, and we thus rule out ordinary neutrino-driven $^{56}$Ni-powered core-collapse SNe as the only power source for LFBOT phenomena. Below, we discuss three main physical models for LFBOTs. 

(i) \citet{Khatami2024ApJ...972..140K} propose that LFBOT UVOIR light curves can be modeled as SNe that explode within dense CSM shells, leading to a shock-breakout flash that produces the initial fast rise to peak emission and a subsequent shock-cooling tail. As the blackbody temperature of \AT\, does not cool beyond $\sim18,000\,$K (for $\delta t\lesssim 55\,$ rest-frame days; see Fig.~\ref{fig:BB_fits}), we note that this pure CSM interaction interpretation would require very long lived, continuous interaction to maintain such temperatures. Additionally, the inferred blackbody expansion velocities of up to 0.3\,c imply a compact-object power source with a large enough gravitational well to be able to launch such fast outflows. But instead, if the outflows are interacting with a pre-existing CSM, the inferred velocity ($dR/dt$) is reduced to a few $10^{-2}-10^{-3}\,$c. However, more importantly, the CSM interaction model struggles to reproduce the X-ray nonthermal spectrum and rapid variability discussed in Paper II.  Thus, we disfavor a pure CSM interaction model for LFBOT phenomena, though we note that CSM interaction could play a role in addition to accretion onto a compact object (e.g., to produce the observed radio emission). We thus consider two alternative models that involve compact-object accretion.  

(ii) Following \citet{tsuna2025}, we consider the production of an LFBOT via a stripped-envelope SN (Type Ibc) that produces a compact object with a kick that is fortuitously aligned such that the compact object can disrupt and accrete its main-sequence companion at a super-Eddington rate. The LFBOT is a consequence of the accretion and occurs with some time delay after the SN. Generally, if the compact object is a BH, super-Eddington accretion could provide the $10^{51}\,$erg radiated by \AT\, (an NS falls just short of producing the  energetics of \AT\, but could be enough for AT\,2018cow; see \citealt{tsuna2025}, their Eqs.~16 and 37) and generate LFBOT inferred asymmetric geometry and multicomponent outflows. \citet{tsuna2025} find that their models are able to reproduce rates consistent with LFBOT observational rates as well as typical LFBOT features including the fast ($\sim0.2$c) outflows originating from the super-Eddington accretion winds, spectra that evolve from featureless to presenting weak H and He lines at $\sim20\,$d with widths of a few $10^{3}\,\rm{km\,s}^{-1}$ and small $L_{\rm{H\alpha}}/L_{\rm{rad}}<1\%$, the evolution of the ratio $L_{\rm X}/L_{\rm UVOIR}$ from $<<1$ to $\sim1$ over the same timescale due to X-ray ionization of the ejecta, and the steep radio-inferred CSM density profile at $R>10^{16}\,$cm likely due to He-star's mass loss within the last few centuries before explosion. 
If the CSM is dense enough, it could naturally produce a dust echo emitting the observed NIR excess as discussed in \S\ref{SubSec:NIRDustEcho}, though free-free opacity effects (\S\ref{SubSec:NIRfreefree}) could also produce this emission component. 

Observationally, the challenge for this model is avoiding the detection of the SN. If there were a Type Ibc SN that occurred just before the LFBOT, the SN would have to be significantly underluminous with a short delay time before the LFBOT in order for the SN to remain undetected. Nothing suggesting the presence of an SN has been detected in an LFBOT thus far. The SN may also dominate the light curve once the FBOT emission fades significantly, though observations of AT\,2018cow at 2--4 yr after discovery showed a bright ($L\approx 10^{39}\,\rm{erg\,s}^{-1}$), blue ($\rm{F336W} - \rm{F555W} = - 1.3$\,mag) source interpreted as a remnant accretion disk around a BH \citep{Sun2022MNRAS.512L..66S, Sun2023MNRAS.519.3785S,Chen2023b}, which could potentially prevent detection of the SN light-curve component. 
We would also expect typical Type Ibc SN spectral features to be produced (e.g.,~Ca, O), which have yet to be observed in LFBOTs.

(iii) \citet{metzger2022aj} model LFBOTs as emission from the tidal disruption of a WR star with a compact-object binary companion (either NS or stellar-mass BH) and subsequent super-Eddington accretion. The optical light curve is produced by some combination of reprocessed X-rays produced by the central accretion disk and shock interaction between the WR ejecta and CSM from prior mass-loss events. Observed X-rays would consist of the fraction that were not reprocessed by the intervening material. This model broadly produces phenomena consistent with AT\,2018cow, and thus \AT, including a viscous accretion timescale of less than a few days, matching with LFBOT peak timescales; energy budget from super-Eddington accretion of $\sim10^{51}\,\rm{erg}$; generation of both fast polar (0.1\,c) and slow equatorial (few $10^4\,\rm{km\,s}^{-1}$) outflows; $L_{\rm{engine}}\propto t^{-2.1}$ for an accretion efficiency of $\eta\approx 0.01$ (reasonable for super-Eddington accretion onto a magnetar or BH) which is slightly shallower than observed in \AT\, ($L_{\rm{engine}}\propto t^{-3.4}$; see Fig.~\ref{fig:Lengine}) but more consistent with the $L_{\rm{engine}}$ evolution of AT\,2018cow ($\propto t^{-1.9}$ \citealt{Margutti19}); low $^{56}$Ni abundance ($<10^{-2}\,\rm{M}_\odot$) in the disk outflows consistent with AT\,2018cow light-curve modeling \citep{perley2019mnras}; rough agreement with the observed optical light curve of AT\,2018cow where the early-time emission is reprocessed X-rays and the resulting X-ray luminosity is suppressed due to this absorption causing $L_X/L_{\rm{UVOIR}}<1$ until $\tau_X\approx 1$ at $\sim20\,$d; and a density of $n\approx 10^{5}\,\rm{cm}^{-3}$ combined with a steepening of the density profile from $n\propto r^{-2}$ to $\propto r^{-3}$ at $r\approx 10^{16}\,\rm{cm}$ in the remnant circumbinary disk which is produced by WR mass loss prior to the explosion and is consistent with radio-inferred densities. 
CSM formed by this mass loss would provide a natural environment for a NIR dust echo to be produced as discussed in \S\ref{SubSec:NIRDustEcho}.
\citet{metzger2022aj} further predict a potential flattening of the light curve at late times ($\gtrsim100\,$d) due to the compact-object accretion rate approaching the Eddington limit, which was observed in AT\,2018cow \citep{Sun2022MNRAS.512L..66S,Sun2023MNRAS.519.3785S,Chen2023b}. Assuming \AT\, had similar luminosities and light-curve evolution to AT\,2018cow (see Fig.~4 of \citealt{Chen2023b}), \textit{HST} could have observed such behavior up to $\sim100$\,d. 

This model predicts an LFBOT continuum where slower evolving transients reach higher luminosities (see Fig.~3 of \citealt{metzger2022aj}), as increasing the system mass (i.e., $M_{\rm BH}$ and the WR mass $M_{\rm WR}$) can lead to both a longer viscous timescale and peak accretion rate, thus flattening and increasing the peak of the light curve. This continuum is broadly consistent with the properties of \AT\, (and AT\,2024puz) compared to other LFBOTs. 

\,\,\,\,\,\,\,\,\,\, \,\,\,\,\, \,\,\,\,\,\,\,\,\,\, \,\,\,\,\,  *  \,\,\,\,\,\,\,\,\,\, \,\,\,\,\,          *       \,\,\,\,\,\,\,\,\,\, \,\,\,\,\,            *

To conclude this subsection, overall the models that are the most successful in accounting for the panchromatic properties of LFBOTs share the common ingredient of a central engine in the form of a super-Eddington accretion disk around a compact object and invoke shocks between outflows launched by the accretion disk and/or between disk outflows and pre-existing CSM as a way to thermalize some of the energy released by the central engine. At the time of writing, the astrophysical origin and nature of the compact object is an open question.

\section{Summary and Conclusions}\label{Sec:Conclusions}

We have presented an extensive UVOIR photometric (Fig.~\ref{fig:lightcurve}) and spectroscopic (Figs.~\ref{fig:optical_spec}, \ref{fig:NIR_spec}) observational campaign for the third LFBOT to be well-sampled in this wavelength range during the time period $\delta t = 0.1$--97\,d. We summarize our findings as follows.

\begin{itemize}
    \item \AT\, is the most luminous LFBOT discovered to date (both at UV wavelengths and bolometrically; see Fig.~\ref{fig:UVcombined}), reaching $L_{\rm{pk}}\approx (2-4)\times 10^{45}\,\rm{erg\,s^{-1}}$ at peak (which is $\sim 5$--10 times larger than the prototypical event of this class, AT\,2018cow). The UV-optical spectrum is dominated by a thermal continuum at all times (Fig.~\ref{fig:optical_spec}).  From our blackbody fits (Fig.~\ref{fig:BB_fits}), we infer $T >30{,}000$\,K and blackbody ``expansion velocities'' of $0.2-0.3c$ at peak --- slightly larger than (but comparable to) the mildly relativistic expansion of the optical photosphere inferred for AT\,2018cow. Similarly to AT\,2018cow, \AT\, maintains a high temperature ($\gtrsim 20{,}000$\,K) throughout our monitoring. We infer an initial $R_{\rm{BB}}\approx 2\times10^{15}$\,cm. Unlike ordinary SNe, $R_{\rm{BB}}$ shows  a brief phase of expansion in the first few days before decreasing monotonically. The post-peak bolometric luminosity shows rapid fading $\propto t^{-3.4}$.
    \item The rise time $t_{\rm{rise}}$ of \AT\, is among the longest of known classical LFBOTs (Fig.~\ref{fig:UVcombined}).  Assuming $t_{\rm{rise}}=4\,$d reflects the diffusion timescale of radiation from a central source (we note this is an upper limit), we infer an ejecta mass $M_{\rm{ej}}\lesssim(1-2)$\,M$_\odot$, which is larger than inferred for AT\,2018cow, but consistent with the longer rise time of \AT\, to peak but similarly mildly relativistic initial blackbody expansion velocity.
    \item Over the first $\sim 45\,$d, \AT\, radiated $>10^{51}\,$erg (Table \ref{tab:E_rad}), more than an order of magnitude above that radiated by AT\,2018cow in a similar timescale, and a value only matched by the most luminous stellar explosions. The large $E_{\rm{rad}}$ rules out ordinary neutrino-driven SNe and requires additional sources of energy. Among these, we favor super-Eddington accretion-powered systems harboring a compact object (most likely a BH).  the other two LFBOTs with existing spectral sequences (CSS161010, \citealt{Gutierrez2024}; AT\,2018cow, \citealt{Margutti19, perley2019mnras}), the spectra of \AT\, are entirely featureless for weeks post discovery (Fig.~\ref{fig:optical_spec}). This is a hallmark observational feature of LFBOTs that sets them apart from SNe. Interestingly, this observational trait is also seen in the new class of ``featureless TDEs'' \citep{Andreoni2023Natur.613E...6A,Hammerstein2023a, Yao2023ApJ...955L...6Y}, with which LFBOTs share the presence of a central source of high-energy emission. By analogy with featureless TDEs, we suggest that featureless spectra might result from persistent ionization of the fast-expanding ejecta. 
    \item At $\delta t > 35\,$d, we confidently detect faint (EW $\lesssim10\,$\AA; Fig.~\ref{fig:Line_params}) spectral features of H and He with two kinematically separate velocity components centered at $0\,\rm{km\,s}^{-1}$  and $-6400\,\rm{km\,s}^{-1}$ with FWHM $\approx 2000\,\rm{km\,s^{-1}}$ (Fig.~\ref{fig:optical_lines}). A prominently blueshifted component was detected before in CSS\,161010 \citep{Gutierrez2024} (Fig.~\ref{fig:optical_halpha}). The line profiles indicate a clear deviation from spherical symmetry. We note that as in AT\,2018cow, the spectral features emerge when $L_X\approx L_{\rm{UVOIR}}$ (Fig.~\ref{fig:Lx/Lbol}). 
    \item These line profiles imply the presence of multiple outflow components --- namely, one that is fast and polar, and another that is slower and equatorial. Super-Eddington accretion disks provide a natural explanation for this structure \citep{Yoshioka24}.  
    \item While overall the UV-optical SED at each epoch is well fit by a blackbody spectrum, between $20-30\,$d we measure a NIR excess of emission, with a power-law spectrum $F_{\rm \nu,NIR}\propto \nu^{-0.3}$ at 30\,d (Fig.~\ref{Sec:NIRexcess}). The presence of NIR excess emission is similar to AT\,2018cow  (\citealt{perley2019mnras,Margutti19,chen2023a}) and AT\,2024puz (\citealt{Somalwar2025}).  
    The extrapolation of the radio SED of \AT\, to NIR wavelengths significantly underpredicts the observed flux, implying that the NIR is a distinct emission component. 
    \item We consider two models to explain the NIR excess: (i) reprocessing of early UV emission by pre-existing dust with density $n_0\approx 4\times 10^7\,\rm{cm}^{-3}$ at $r_0=10^{16}\,\rm{cm}$  (\S\ref{SubSec:NIRDustEcho}; following \citealt{Metzger&Perley2023, tuna2025}), which is consistent with the radio-inferred density from Paper II ($\sim 10^5\,\rm{cm^{-3}}$) with large deviation from equipartition; and (ii) free-free scattering occurring in the extended medium above the optical photosphere (\S\ref{SubSec:NIRfreefree}; following \citealt{Chen_Shen_2024,Roth16,lu2020mnras, Somalwar2025}) with density profile $\rho(r)=\rho_0(r_0/r)^{1.3}$. Both models are able to produce the observed $L_{\rm NIR}=(1.9\pm 0.3) \times 10^{41}\,\rm{erg\,s}^{-1}$ at the time of appearance (30\,d). The free-free model requires a density $\rho_0\approx 7\times10^{-16}\,\rm{g\,cm}^{-3}$ at $r_0\approx10^{16}\,$cm, which is $\sim10^4$ times the radio-derived CSM density; thus, in this model we would expect the NIR to originate from CSM separate from the radio-emitting region.
    However, neither model provides a natural explanation for the roughly constant soft X-ray to NIR luminosity ratio of $L_{\rm X}/L_{\rm NIR}\approx 0.5$ (Fig.~\ref{fig:Lx/Lbol}).     
\end{itemize}

While the presence of a compact-object central engine is a feature of LFBOT models that successfully reproduce LFBOT phenomenology across the electromagnetic spectrum, the nature of this object is unknown. Circumstantial evidence such as the extreme radiated energy of \AT\, ($E_{rad}\approx 10^{51}\,$erg) and mildly relativistic velocities inferred in LFBOT radio and optical emission suggest stellar-mass BHs, which would make LFBOTs highly super-Eddington, and thus valuable probes of this accretion regime. Progress relies on increasing the small sample of well-studied objects.

As LFBOTs are extremely UV-luminous transients, future UV missions such as ULTRASAT \citep{Sagiv2014AJ....147...79S} and UVEX \citep{Kulkarni2021arXiv211115608K} will be instrumental for discovering and characterizing these rare transients out to larger volumes and at earlier times. Tens of LFBOTs per year will be discovered by these surveys. Discovering more LFBOTs pre-peak will also allow better sampling of color evolution from pre- to post-rise, potentially revealing the initial reddened colors indicative of sublimation of dusty CSM predicted by \citet{Metzger&Perley2023}. 
At the same time, on the follow-up side, higher-cadence NIR monitoring coordinated with X-ray observations to later epochs is also needed to understand whether the X-ray/NIR correlation with time observed in AT\,2018cow (Fig.~5 of \citealt{chen2023a}) is distinctive to LFBOTs and thus requires a model that connects the two emission components. 
Mid-IR spectroscopy with \textit{JWST} would also better constrain the SED peak of the observed NIR excess and simultaneously be sensitive to dust features (e.g., the $\sim9\,\mu$m feature indicating silicate dust composition). 
We conclude by emphasizing that only two LFBOTs (three if including AT\,2024puz) to date have extensive multiwavelength datasets; thus, future observations are required to probe the diversity of this class and put meaningful population-level constraints on the intrinsic nature of these intriguing objects.

\vspace{5mm}

\facilities{Swift (XRT and UVOT), Keck, Lick, ATLAS, Supra Solem, REM, Gemini, Thacher, SALT, LDT, GALEX}
\software{Astropy \citep{astropy:2013,astropy:2018, astropy:2022}, NumPy \citep{harris2020array}, synphot \citep{2018ascl.soft11001S}, Matplotlib \citep{Hunter2007matplotlib}, photutils \citep{photutils:v.1.11.0}, HOTPANTS \citep{Becker15}, photpipe \citep{Rest05}, DRAGONS \citep{DRAGONS,DRAGONS:zendo}, SWarp \citep{swarp}, DoPhot \citep{Schechter93}, UCSC Spectral Pipeline \citep{Siebert20}, LPipe \citep{Perley2019PASP}, PySALT \citep{Crawford_pysalt_2010}, PypeIt \citep{Prochaska2020_pypeit, Prochaska_pypeit_zenodo}, Astroalign \citep{Beroiz20}}

\section*{Acknowledgments}

Some of the data presented herein were obtained at Keck Observatory, which is a private 501(c)3 nonprofit organization operated as a scientific partnership among the California Institute of Technology, the University of California, and the National Aeronautics and Space Administration (NASA). The Observatory was made possible by the generous financial support of the W. M. Keck Foundation. 

The authors wish to recognize and acknowledge the very significant cultural role and reverence that the summit of Maunakea has always had within the Native Hawaiian community. We are most fortunate to have the opportunity to conduct observations from this mountain.

Some of the LRIS data presented in this work were the result of cooperative data-sharing agreements between the classical and ToO programs of our group with another Keck ToO proposal (Keck 2024B proposal U042; PI: S. Valenti) and classical observers (D. Perley and I. Caiazzo).

Observations used in this work were obtained in part at the international Gemini Observatory, a program of NSF's NOIRLab, which is managed by the Association of Universities for Research in Astronomy (AURA, Inc.) under a cooperative agreement with the U.S. National Science Foundation (NSF) on behalf of the Gemini Observatory partnership: the NSF (United States), National Research Council (Canada), Agencia Nacional de Investigaci\'{o}n y Desarrollo (Chile), Ministerio de Ciencia, Tecnolog\'{i}a e Innovaci\'{o}n (Argentina), Minist\'{e}rio da Ci\^{e}ncia, Tecnologia, Inova\c{c}\~{o}es e Comunica\c{c}\~{o}es (Brazil), and Korea Astronomy and Space Science Institute (Republic of Korea).

A major upgrade of the Kast spectrograph on the Shane 3 m telescope at Lick Observatory, led by Brad Holden, was made possible through gifts from the Heising-Simons Foundation, William and Marina Kast, and the University of California Observatories. Research
at Lick Observatory is partially supported by a generous gift from Google.

These results made use of the Lowell Discovery Telescope (LDT) at Lowell Observatory. Lowell is a private, nonprofit institution dedicated to astrophysical research and public appreciation of astronomy and operates the LDT in partnership with Boston University, the University of Maryland, the University of Toledo, Northern Arizona University and Yale University. The upgrade of the DeVeny optical spectrograph has been funded by a generous grant from John and Ginger Giovale and by a grant from the Mt.~Cuba Astronomical Foundation.

The Southern African Large Telescope (SALT) observations presented here were taken under Rutgers University program 2024-1-MLT-003 (PI: S. W. Jha). Research on astrophysical transients at Rutgers University is supported by NSF grant AST-2407567.

This research has made use of data and software provided by the High Energy Astrophysics Science Archive Research Center (HEASARC), which is a service of the Astrophysics Science Division at NASA/GSFC.

This work has made use of data from the Asteroid Terrestrial-impact Last Alert System (ATLAS) project. The 
ATLAS project is primarily funded to search for near-Earth objects through NASA grants NN12AR55G, 80NSSC18K0284, and 80NSSC18K1575; byproducts of the NEO search include images and catalogs from the survey area. This work was partially funded by Kepler/K2 grant J1944/80NSSC19K0112 and HST GO-15889, and STFC grants ST/T000198/1 and ST/S006109/1. The ATLAS science products have been made possible through the contributions of the University of Hawaii Institute for Astronomy, the Queen’s University Belfast, the Space Telescope Science Institute, the South African Astronomical Observatory, and The Millennium Institute of Astrophysics (MAS), Chile.

\texttt{BLAST} makes use of the following software packages: \texttt{AstroPy} \citep{Astropy_2018}, \texttt{NumPy} \citep{Harris_2020}, GHOST \citep{Gagliano_2021}, \texttt{PhotUtils} \citep{Bradley_2024}, \texttt{Astroquery} \citep{Ginsburg_2019}, \texttt{HiPS} \citep{Fernique_2015}, \texttt{DYNESTY} \citep{Speagle_2020},  \texttt{Prospector} \citep{Johnson_2021}, \texttt{sedpy} \citep{Johnson_2021_SEDPY}, SVO Filter Profile Service \citep{Rodrigo_2020}, \texttt{Healpy} \citep{Zonca_2019}, \texttt{Python FSPS} \citep{Johnson_2024}, \texttt{SBI} \citep{Tejero-Cantero_2020sbi}, and \texttt{SBI++} \citep{Wang_2023}. 

R.M. acknowledges support by the National Science
Foundation under award No. AST-2224255, and by NASA under grants 80NSSC22K1587, 80NSSC25K7591 and 80NSSC22K0898. 
A.V.F.’s research group at U.C. Berkeley acknowledges financial                 
assistance from the Christopher R. Redlich Fund, as well                      
as donations from Gary and Cynthia Bengier, Clark                             
and Sharon Winslow, Alan Eustace and Kathy Kwan,                              
William Draper, Timothy and Melissa Draper, Briggs                            
and Kathleen Wood, Sanford Robertson (W.Z. 
 is a Bengier-Winslow-Eustace Specialist in Astronomy,                         
 T.G.B. is a Draper-Wood-Robertson Specialist in Astronomy, Y.Y. was a Bengier-Winslow-Robertson Fellow in Astronomy), and many other donors.   
N.L. thanks the LSST-DA Data Science Fellowship Program, which is funded by LSST-DA, the Brinson Foundation, the WoodNext Foundation, and the Research Corporation for Science Advancement Foundation; her participation in the program has benefited this work.
C.D.K. gratefully acknowledges support from the NSF through grant AST-2432037, the HST Guest Observer Program through HST-SNAP-17070 and HST-GO-17706, and from JWST Archival Research through JWST-AR-6241 and JWST-AR-5441.
G.M. acknowledges financial support from the INAF mini-grant "The high-energy view of jets and transient" (Bando Ricerca Fondamentale INAF 2022).

\appendix

\setcounter{table}{0}
\renewcommand{\thetable}{B\arabic{table}}

\begin{figure*}[h]
\centering
\includegraphics[width=\columnwidth]{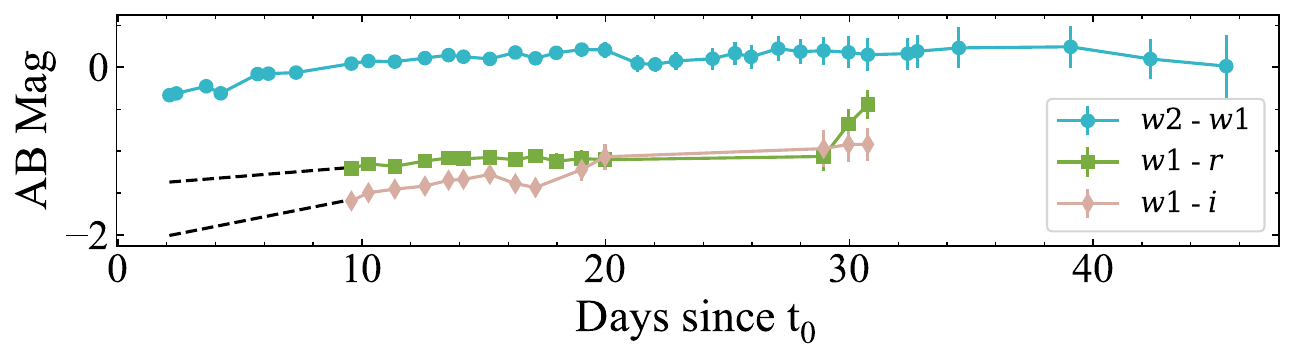}
  \caption{Color evolution of \AT. The black dashed lines are extrapolations of the {$w1$} -- \textit{r} and {$w1$} -- \textit{i} colors to the beginning of the Swift photometry on day +2, which were used to derive the blackbody temperature and radius fits presented in Fig. \ref{fig:BB_fits}. }
  \label{fig:color}
\end{figure*}

\section{AT~2023fhn {Swift} {$w1$} photometry}
\label{AT2023fhn_appendix}
{Swift}-UVOT observed AT\,2023fhn beginning at 2023-04-25 08:24:30 ($\delta t = 15.1\,$d). We report $20.68 \pm 0.08$\,mag as the non-extinction-corrected {$w1$} magnitude. We reduced and corrected this photometry for Galactic extinction using the methods performed for the \AT\, {Swift} UVOT photometry in \S\ref{sec:observations}. MW extinction was corrected using $R_V =3.1$ with the \citet{F99MWExtCorr} model with $A_{V} = 0.078$\,mag and $E(B-V) = 0.025$\,mag. 

\section{Data Tables}

\startlongtable
\begin{deluxetable*}{ccccc}
\tablecaption{ATLAS Optical Photometry (AB Mag)}
\tablewidth{0pt}
\tablehead{
\colhead{UTC Date} &
\colhead{MJD} &
\colhead{Phase\tablenotemark{a}} &
\colhead{Mag} &
\colhead{Filter}
}
\decimalcolnumbers
\startdata
2024-09-24 & 60577.3 & -1.0 & $>$ 18.3\tablenotemark{b} & o \\
2024-09-27 & 60580.0 & 1.7 & 17.54 $\pm$ 0.03 & o \\
2024-10-01 & 60584.9 & 6.6 & 16.88 $\pm$ 0.02 & o \\
2024-10-03 & 60587.0 & 8.7 & 17.20 $\pm$ 0.02 & o \\
2024-10-04 & 60588.0 & 9.7 & 17.33 $\pm$ 0.02 & o \\
2024-10-08 & 60591.5 & 13.2 & 18.20 $\pm$ 0.04 & o \\
2024-10-09 & 60593.0 & 14.6 & 18.44 $\pm$ 0.06 & o \\
2024-10-10 & 60593.1 & 14.8 & 18.6 $\pm$ 0.2 & o \\
2024-10-11 & 60594.6 & 16.3 & 18.61 $\pm$ 0.06 & o \\
2024-10-12 & 60595.9 & 17.6 & 19.1 $\pm$ 0.2 & o \\
2024-10-15 & 60598.4 & 20.1 & 19.5 $\pm$ 0.2 & o \\
2024-10-17 & 60600.1 & 21.7 & 18.9 $\pm$ 0.3 & o \\
2024-10-19 & 60602.3 & 23.9 & 19.5 $\pm$ 0.3 & o \\
2024-10-22 & 60605.3 & 26.9 & $>$ 19.6\tablenotemark{b} & o \\
2024-10-23 & 60606.2 & 27.9 & $>$ 19.5\tablenotemark{b} & o \\
2024-09-28 & 60581.3 & 3.0 & 16.48 $\pm$ 0.01 & c \\
2024-09-29 & 60582.2 & 3.9 & 16.30 $\pm$ 0.01 & c \\
2024-10-03 & 60586.2 & 7.9 & 16.65 $\pm$ 0.01 & c \\
2024-10-10 & 60593.2 & 14.9 & 18.28 $\pm$ 0.04 & c \\
2024-10-23 & 60606.9 & 28.5 & 20.3 $\pm$ 0.2 & c \\
2024-10-24 & 60608.0 & 29.6 & $>$ 20.1\tablenotemark{b} & c \\
2024-10-25 & 60608.0 & 29.7 & $>$ 20.1\tablenotemark{b} & c \\
\enddata
\tablenotetext{a}{Relative to $t_0$ (MJD 60578.3).}
\tablenotetext{b}{Upper limit.}
\label{tab:atlas_phot_all}
\end{deluxetable*}

\begin{deluxetable*}{ccccc}[h]
\tablecaption{Public ZTF AB Optical Photometry  \label{tab:optical_phot_all}}
\tablewidth{0pt}
\tablehead{
\colhead{UTC Date} & 
\colhead{MJD} &
\colhead{Phase\tablenotemark{a}} &
\colhead{Mag} &
\colhead{Filter}
}
\decimalcolnumbers
\startdata
2024-09-25 & 60578.44 & 0.11 & 20.7 $\pm$ 0.3 & \textit{g} \\
2024-09-26 & 60579.37 & 1.0 & 17.46 $\pm$ 0.05 & \textit{g} \\
\enddata
\tablenotetext{a}{Relative to $t_0$ (MJD 60578.3).}
\label{tab:ZTF_public_phot}
\end{deluxetable*}

\startlongtable
\begin{deluxetable*}{cccccc}
\tablecaption{Swift UVOT AB Optical Photometry }
\tablewidth{0pt}
\tablehead{
\colhead{UTC Date} &
\colhead{MJD} &
\colhead{Phase\tablenotemark{a}} &
\colhead{$U$} &
\colhead{$B$} &
\colhead{$V$}
}
\decimalcolnumbers
\startdata
2024-09-27 & 60580.40 & 2.1 & 16.01 $\pm$ 0.05 & 16.46 $\pm$ 0.06\\
2024-09-27 & 60580.50 & 2.2 & 16.03 $\pm$ 0.05 & 16.43 $\pm$ 0.06 & 16.9 $\pm$ 0.1\\
2024-09-27 & 60580.50 & 2.2 & 16.02 $\pm$ 0.05 & 16.45 $\pm$ 0.07 &  \\
2024-09-27 & 60580.60 & 2.3 & 15.99 $\pm$ 0.04 & 16.43 $\pm$ 0.06 &  \\
2024-09-27 & 60580.60 & 2.3 & 16.01 $\pm$ 0.05 & 16.37 $\pm$ 0.06 &  \\
2024-09-27 & 60580.70 & 2.4 & 15.99 $\pm$ 0.04 & 16.36 $\pm$ 0.06 & 16.8 $\pm$ 0.1\\
2024-09-28 & 60581.80 & 3.5 & 15.65 $\pm$ 0.04 & 16.04 $\pm$ 0.05\\
2024-09-29 & 60582.00 & 3.7 & 15.60 $\pm$ 0.04 & 16.01 $\pm$ 0.05\\
2024-09-29 & 60582.40 & 4.1 & 15.52 $\pm$ 0.04 & 15.99 $\pm$ 0.06 &  \\
2024-09-29 & 60582.40 & 4.1 & 15.56 $\pm$ 0.04 &   &  \\
2024-09-29 & 60582.50 & 4.2 & 15.59 $\pm$ 0.04 & 15.98 $\pm$ 0.05\\
2024-09-30 & 60583.70 & 5.4 & 15.60 $\pm$ 0.04 & 16.12 $\pm$ 0.08\\
2024-10-01 & 60584.00 & 5.7 & 15.63 $\pm$ 0.05 & 16.03 $\pm$ 0.06 & 16.5 $\pm$ 0.1\\
2024-10-01 & 60584.10 & 5.8 & 15.63 $\pm$ 0.05 & 15.94 $\pm$ 0.07 & 16.3 $\pm$ 0.1\\
2024-10-01 & 60584.30 & 6.0 & 15.67 $\pm$ 0.06 & 16.09 $\pm$ 0.08 & 16.3 $\pm$ 0.2\\
2024-10-01 & 60584.70 & 6.4 & 15.63 $\pm$ 0.04 & 16.11 $\pm$ 0.05 & 16.52 $\pm$ 0.09\\
2024-10-02 & 60585.60 & 7.3 & 15.84 $\pm$ 0.04 & 16.22 $\pm$ 0.05 & 16.66 $\pm$ 0.09\\
2024-10-04 & 60587.80 & 9.5 & 16.26 $\pm$ 0.05 & 16.77 $\pm$ 0.06 & 17.1 $\pm$ 0.1\\
2024-10-05 & 60588.40 & 10.1 & 16.44 $\pm$ 0.05 & 16.88 $\pm$ 0.07 & 17.2 $\pm$ 0.1\\
2024-10-07 & 60590.40 & 12.1 & 16.83 $\pm$ 0.06 & 17.28 $\pm$ 0.08 & 17.7 $\pm$ 0.2\\
2024-10-07 & 60590.50 & 12.2 & 16.88 $\pm$ 0.04 & 17.49 $\pm$ 0.06 & 17.9 $\pm$ 0.2\\
2024-10-07 & 60590.70 & 12.4 & 16.89 $\pm$ 0.04 & 17.37 $\pm$ 0.05 & 17.8 $\pm$ 0.1\\
2024-10-11 & 60594.40 & 16.1 & 18.3 $\pm$ 0.2\\
2024-10-11 & 60594.60 & 16.3 & 17.74 $\pm$ 0.07 & 18.3 $\pm$ 0.1\\
2024-10-12 & 60595.40 & 17.1 & 18.01 $\pm$ 0.09 & 18.2 $\pm$ 0.1\\
2024-10-13 & 60596.80 & 18.5 & 18.3 $\pm$ 0.2\\
2024-10-14 & 60597.30 & 19.0 & 18.3 $\pm$ 0.1 & 18.7 $\pm$ 0.2\\
2024-10-16 & 60599.60 & 21.3 & 18.6 $\pm$ 0.1\\
2024-10-17 & 60600.30 & 22.0 & 18.8 $\pm$ 0.1\\
2024-10-17 & 60600.40 & 22.1 & 19.0 $\pm$ 0.3\\
2024-10-18 & 60601.20 & 22.9 & 18.9 $\pm$ 0.1\\
2024-10-19 & 60602.40 & 24.1 & 19.4 $\pm$ 0.2\\
2024-10-19 & 60602.70 & 24.4 & 19.2 $\pm$ 0.2\\
2024-10-20 & 60603.50 & 25.2 & 19.4 $\pm$ 0.4\\
2024-10-20 & 60603.60 & 25.3 & 19.2 $\pm$ 0.2\\
2024-10-21 & 60604.80 & 26.5 & 19.2 $\pm$ 0.2\\
2024-10-22 & 60605.40 & 27.1 & 19.7 $\pm$ 0.2\\
2024-10-23 & 60606.30 & 28.0 & 20.1 $\pm$ 0.3\\
2024-10-24 & 60607.10 & 28.8 & 19.4 $\pm$ 0.2\\
2024-10-24 & 60607.30 & 29.0 & 19.7 $\pm$ 0.3\\
2024-10-25 & 60608.60 & 30.3 & 19.4 $\pm$ 0.2\\
2024-10-26 & 60609.00 & 30.7 & 19.3 $\pm$ 0.2\\
2024-10-27 & 60610.70 & 32.4 & 19.9 $\pm$ 0.3\\
2024-10-28 & 60611.90 & 33.6 & 20.2 $\pm$ 0.3\\
2024-10-29 & 60612.80 & 34.5 & 20.2 $\pm$ 0.6 & 19.4 $\pm$ 0.7\\
2024-11-07 & 60621.70 & 43.4 & 20.7 $\pm$ 0.3\\
2024-11-11 & 60625.70 & 47.4 & 21.0 $\pm$ 0.4\\
\enddata
\tablenotetext{a}{Relative to $t_0$ (MJD 60578.3).}
\label{tab:swift_phot_UBV}
\end{deluxetable*}

\startlongtable
\begin{deluxetable*}{cccccc}
\tablecaption{Swift UVOT AB UV Photometry }
\tablewidth{0pt}
\tablehead{
\colhead{UTC Date} &
\colhead{MJD} &
\colhead{Phase\tablenotemark{a}} &
\colhead{{$w1$}} &
\colhead{{$w2$}} &
\colhead{{$m2$}}
}
\decimalcolnumbers
\startdata
2024-09-27 & 60580.4 & 2.1 & 15.74 $\pm$ 0.04 & 15.52 $\pm$ 0.04\\
2024-09-27 & 60580.5 & 2.2 &   & 15.76 $\pm$ 0.04 & 15.49 $\pm$ 0.04\\
2024-09-27 & 60580.5 & 2.2 &   & 15.68 $\pm$ 0.04 &  \\
2024-09-27 & 60580.6 & 2.3 &   & 15.72 $\pm$ 0.04 & 15.39 $\pm$ 0.05\\
2024-09-27 & 60580.6 & 2.3 &   & 15.63 $\pm$ 0.04 & 15.42 $\pm$ 0.04\\
2024-09-27 & 60580.7 & 2.4 & 15.48 $\pm$ 0.04 & 15.64 $\pm$ 0.04 & 15.39 $\pm$ 0.04\\
2024-09-28 & 60581.8 & 3.5 & 15.29 $\pm$ 0.04\\
2024-09-28 & 60581.9 & 3.6 & 15.33 $\pm$ 0.04\\
2024-09-29 & 60582.0 & 3.7 & 15.32 $\pm$ 0.04 & 15.17 $\pm$ 0.04\\
2024-09-29 & 60582.4 & 4.1 & 15.29 $\pm$ 0.04\\
2024-09-29 & 60582.5 & 4.2 & 15.29 $\pm$ 0.04 & 15.14 $\pm$ 0.04\\
2024-09-30 & 60583.7 & 5.4 & 15.36 $\pm$ 0.04\\
2024-10-01 & 60584.0 & 5.7 & 15.27 $\pm$ 0.04 & 15.41 $\pm$ 0.04 & 15.42 $\pm$ 0.04\\
2024-10-01 & 60584.1 & 5.8 & 15.31 $\pm$ 0.04 & 15.41 $\pm$ 0.05 & 15.39 $\pm$ 0.04\\
2024-10-01 & 60584.3 & 6.0 & 15.33 $\pm$ 0.05 & 15.42 $\pm$ 0.04\\
2024-10-01 & 60584.7 & 6.4 & 15.38 $\pm$ 0.04 & 15.47 $\pm$ 0.04 & 15.49 $\pm$ 0.04\\
2024-10-02 & 60585.6 & 7.3 & 15.53 $\pm$ 0.04 & 15.60 $\pm$ 0.04 & 15.64 $\pm$ 0.04\\
2024-10-04 & 60587.8 & 9.5 & 16.03 $\pm$ 0.04 & 16.09 $\pm$ 0.04 & 16.18 $\pm$ 0.04\\
2024-10-05 & 60588.4 & 10.1 & 16.23 $\pm$ 0.04 & 16.23 $\pm$ 0.04 & 16.37 $\pm$ 0.04\\
2024-10-07 & 60590.4 & 12.1 & 16.79 $\pm$ 0.05 & 16.78 $\pm$ 0.05 & 16.95 $\pm$ 0.05\\
2024-10-07 & 60590.5 & 12.2 & 16.80 $\pm$ 0.04 & 16.80 $\pm$ 0.04 & 17.00 $\pm$ 0.04\\
2024-10-07 & 60590.7 & 12.4 & 16.74 $\pm$ 0.04 & 16.84 $\pm$ 0.04 & 16.95 $\pm$ 0.04\\
2024-10-09 & 60592.5 & 14.2 & 17.16 $\pm$ 0.05\\
2024-10-10 & 60593.5 & 15.2 & 17.50 $\pm$ 0.05 & 17.45 $\pm$ 0.06 & 17.62 $\pm$ 0.05\\
2024-10-11 & 60594.6 & 16.3 & 17.83 $\pm$ 0.05 & 17.69 $\pm$ 0.06 & 17.93 $\pm$ 0.05\\
2024-10-12 & 60595.4 & 17.1 & 17.87 $\pm$ 0.05 & 17.82 $\pm$ 0.06 & 17.97 $\pm$ 0.05\\
2024-10-13 & 60596.3 & 18.0 & 18.14 $\pm$ 0.06\\
2024-10-14 & 60597.3 & 19.0 & 18.21 $\pm$ 0.06 & 18.11 $\pm$ 0.07 & 18.40 $\pm$ 0.06\\
2024-10-16 & 60599.6 & 21.3 & 18.61 $\pm$ 0.07 & 18.68 $\pm$ 0.09 & 18.76 $\pm$ 0.07\\
2024-10-17 & 60600.3 & 22.0 & 18.99 $\pm$ 0.08 & 19.1 $\pm$ 0.1 & 19.36 $\pm$ 0.09\\
2024-10-18 & 60601.2 & 22.9 & 18.93 $\pm$ 0.07 & 18.9 $\pm$ 0.1 & 19.04 $\pm$ 0.08\\
2024-10-19 & 60602.7 & 24.4 & 19.22 $\pm$ 0.08 & 19.1 $\pm$ 0.1 & 19.4 $\pm$ 0.1\\
2024-10-20 & 60603.6 & 25.3 & 19.25 $\pm$ 0.09 & 19.2 $\pm$ 0.1 & 19.25 $\pm$ 0.09\\
2024-10-22 & 60605.4 & 27.1 & 19.48 $\pm$ 0.09 & 19.6 $\pm$ 0.1\\
2024-10-23 & 60606.3 & 28.0 & 19.7 $\pm$ 0.1 & 19.5 $\pm$ 0.1 & 20.0 $\pm$ 0.1\\
2024-10-24 & 60607.1 & 28.8 & 19.8 $\pm$ 0.1 & 19.7 $\pm$ 0.2 & 20.0 $\pm$ 0.1\\
2024-10-25 & 60608.6 & 30.3 & 19.9 $\pm$ 0.2 & 19.7 $\pm$ 0.1 & 19.7 $\pm$ 0.1\\
2024-10-26 & 60609.0 & 30.7 & 19.6 $\pm$ 0.1 & 19.7 $\pm$ 0.1 & 19.9 $\pm$ 0.1\\
2024-10-27 & 60610.7 & 32.4 & 20.2 $\pm$ 0.1 & 20.1 $\pm$ 0.1\\
2024-10-28 & 60611.1 & 32.8 & 20.3 $\pm$ 0.1 & 20.4 $\pm$ 0.2\\
2024-10-29 & 60612.8 & 34.5 & 20.2 $\pm$ 0.1 & 20.1 $\pm$ 0.2 & 20.5 $\pm$ 0.1\\
2024-11-06 & 60620.6 & 42.3 & 20.9 $\pm$ 0.1\\
2024-11-07 & 60621.7 & 43.4 & 21.0 $\pm$ 0.2 & 21.0 $\pm$ 0.1\\
2024-11-09 & 60623.8 & 45.5 & 21.0 $\pm$ 0.2\\
2024-11-10 & 60624.7 & 46.4 & 21.3 $\pm$ 0.2\\
2024-11-11 & 60625.7 & 47.4 & 20.8 $\pm$ 0.2 & 20.9 $\pm$ 0.1\\
2024-11-18 & 60632.8 & 54.5 & 21.3 $\pm$ 0.2\\
2024-11-19 & 60633.7 & 55.4 & 21.2 $\pm$ 0.3\\
2024-11-22 & 60636.1 & 57.8 & 21.9 $\pm$ 0.3\\
2024-11-22 & 60636.5 & 58.2 & 21.3 $\pm$ 0.2 & 21.9 $\pm$ 0.2\\
2024-11-23 & 60637.4 & 59.1 & 21.6 $\pm$ 0.2\\
2024-11-28 & 60642.3 & 64.0 & 21.5 $\pm$ 0.2 & 21.8 $\pm$ 0.2\\
2024-12-12 & 60656.3 & 78.0 & 22.1 $\pm$ 0.3 & 21.9 $\pm$ 0.2\\
2024-12-31 & 60675.2 & 96.9 & 22.2 $\pm$ 0.3\\
\enddata
\tablenotetext{a}{Relative to $t_0$ (MJD 60578.3).}
\label{tab:swift_phot_UVW}
\end{deluxetable*}

\startlongtable
\begin{deluxetable*}{cccccc}
\tablecaption{Supra Solem Optical Photometry (AB Mag)}
\tablewidth{0pt}
\tablehead{
\colhead{UTC Date} &
\colhead{MJD} &
\colhead{Phase\tablenotemark{a}} &
\colhead{\textit{g}} &
\colhead{\textit{r}} &
\colhead{\textit{i}}
}
\decimalcolnumbers
\startdata
2024-10-03 & 60586.4 & 8.1 &  & 16.94 $\pm$ 0.06 &  17.5 $\pm$ 0.1 \\
2024-10-04 & 60587.4 &  9.0 & & 17.02 $\pm$ 0.07 &  17.8 $\pm$ 0.1 \\
2024-10-05 & 60588.4 & 10.0 &  & 17.65 $\pm$ 0.08 &  17.9 $\pm$ 0.2 \\
2024-10-09 & 60592.5 & 14.1 & 17.86 $\pm$ 0.09 &  18.5 $\pm$ 0.1 &   \\
2024-10-11 & 60594.3 & 16.0  & & 18.6 $\pm$ 0.2 &   \\
2024-10-12 & 60595.5 & 17.1 & 18.5 $\pm$ 0.1 & 18.9 $\pm$ 0.2 &   \\
\enddata
\tablenotetext{a}{Relative to $t_0$ (MJD 60578.3).}
 \label{tab:supra_solem_phot_all}
\end{deluxetable*}

\startlongtable
\begin{deluxetable*}{cccccc}
\tablecaption{LCO Optical Photometry (AB Mag)}
\tablewidth{0pt}
\tablehead{
\colhead{UTC Date} &
\colhead{MJD} &
\colhead{Phase\tablenotemark{a}} &
\colhead{$g$} &
\colhead{$r$} &
\colhead{$i$}
}
\decimalcolnumbers
\startdata
2024-10-02 & 60585.9 & 7.5 & 16.41 $\pm$ 0.05 & 16.9 $\pm$ 0.2 & 17.39 $\pm$ 0.06\\
2024-10-03 & 60586.7 & 8.3 & 16.61 $\pm$ 0.01 & 17.01 $\pm$ 0.02 & 17.50 $\pm$ 0.02\\
2024-10-04 & 60587.9 & 9.5 & 16.87 $\pm$ 0.03 & 17.27 $\pm$ 0.03 & 17.67 $\pm$ 0.03\\
2024-10-04 & 60587.9 & 9.6 &  &  17.34 $\pm$ 0.02  & 17.64 $\pm$ 0.04\\
2024-10-05 & 60588.6 & 10.2 & 17.11 $\pm$ 0.02 & 17.47 $\pm$ 0.02 & 17.83 $\pm$ 0.05\\
2024-10-06 & 60589.7 & 11.4 & 17.43 $\pm$ 0.02 & 17.79 $\pm$ 0.02 & 18.01 $\pm$ 0.04\\
2024-10-07 & 60590.9 & 12.6 & 17.69 $\pm$ 0.04 & 17.97 $\pm$ 0.05 & 18.27 $\pm$ 0.06\\
2024-10-10 & 60593.2 & 14.9 & 18.13 $\pm$ 0.03 & 18.43 $\pm$ 0.04 & 18.59 $\pm$ 0.05\\
2024-10-10 & 60593.7 & 15.4 & 18.21 $\pm$ 0.04 & 18.61 $\pm$ 0.04 & 18.83 $\pm$ 0.05\\
2024-10-12 & 60595.4 & 17.1 & 18.51 $\pm$ 0.07 & 18.91 $\pm$ 0.06 & 19.3 $\pm$ 0.1\\
2024-10-15 & 60598.3 & 19.9 & 19.19 $\pm$ 0.07 & 19.47 $\pm$ 0.09 & 19.4 $\pm$ 0.1\\
2024-10-24 & 60607.2 & 28.9 & 20.09 $\pm$ 0.06 & 20.7 $\pm$ 0.1 & 20.6 $\pm$ 0.2\\
2024-10-25 & 60608.3 & 29.9 & 20.07 $\pm$ 0.05 & 20.49 $\pm$ 0.09 & 20.7 $\pm$ 0.1\\
\enddata
\tablenotetext{a}{Relative to $t_0$ (MJD 60578.3).}
 \label{tab:LCO_phot}
\end{deluxetable*}

\startlongtable
\begin{deluxetable*}{cccccccc}
\tablecaption{REM Photometry (AB Mag)}
\tablewidth{0pt}
\tablehead{
\colhead{UTC Date} &
\colhead{MJD} &
\colhead{Phase\tablenotemark{a}} &
\colhead{\textit{g}} &
\colhead{\textit{r}} &
\colhead{\textit{i}} &
\colhead{J} &
\colhead{H}
}
\decimalcolnumbers
\startdata
2024-10-02 & 60585.0 & 6.7 & 16.35 $\pm$ 0.04 & 16.63 $\pm$ 0.04 & 17.02 $\pm$ 0.07 & $>$ 17.2\tablenotemark{b} & $>$ 17.7\tablenotemark{b}\\
2024-10-03 & 60586.0 & 7.7 & 16.60 $\pm$ 0.03 & 16.94 $\pm$ 0.03 & 17.19 $\pm$ 0.06 & $>$ 17.6\tablenotemark{b} & \\
2024-10-12 & 60595.0 & 16.7 &  &  &  &  $>$ 17.5\tablenotemark{b} & \\
\enddata
\tablenotetext{a}{Relative to $t_0$ (MJD 60578.3).}
 \label{tab:REM_phot_all}
\tablenotetext{b}{Upper limit.}
\end{deluxetable*}

\startlongtable
\begin{deluxetable*}{cccccc}
\tablecaption{Gemini NIR Photometry AB Mag}
\tablewidth{0pt}
\tablehead{
\colhead{UTC Date} &
\colhead{MJD} &
\colhead{Phase\tablenotemark{a}} &
\colhead{$J$} &
\colhead{$H$} &
\colhead{$K_s$}
}
\decimalcolnumbers
\startdata
2024-10-05 & 60588.3 & 10.0 & 18.79 $\pm$ 0.04 & 19.21 $\pm$ 0.06 &   \\
2024-10-16 & 60599.2 & 20.9 & 20.47 $\pm$ 0.08 &  &   \\
2024-10-17 & 60600.2 & 21.9 & 20.6 $\pm$ 0.1 &  &   \\
2024-10-25 & 60608.3 & 30.0 & 21.2 $\pm$ 0.2 & 21.4 $\pm$ 0.1 &  21.0 $\pm$ 0.2\\
\enddata
\tablenotetext{a}{Relative to $t_0$ (MJD 60578.3).}
 \label{tab:gemini_phot}
\end{deluxetable*}

\startlongtable
\begin{deluxetable*}{ccccccc}
\tablecaption{Thacher Optical Photometry (AB Mag)}
\tablewidth{0pt}
\tablehead{
\colhead{UTC Date} &
\colhead{MJD} &
\colhead{Phase\tablenotemark{a}} &
\colhead{\textit{g}} &
\colhead{\textit{r}} &
\colhead{\textit{i}} &
\colhead{\textit{z}}
}
\decimalcolnumbers
\startdata
2024-10-04 & 60587.3 & 9.0 & 16.77 $\pm$ 0.02 &  17.16 $\pm$ 0.02 &17.57 $\pm$ 0.04 & 17.57 $\pm$ 0.09\\
2024-10-05 & 60588.3 & 10.0 & 17.07 $\pm$ 0.02 &  17.36 $\pm$ 0.02 & 17.68 $\pm$ 0.03 &  17.74 $\pm$ 0.08\\
2024-10-11 & 60594.2 & 15.9 & 18.4 $\pm$ 0.1 &  &  &     \\
2024-10-13 & 60596.2 & 17.9 & 18.9 $\pm$ 0.1 & 18.95 $\pm$ 0.08 & & 19.5 $\pm$ 0.2\\
\enddata
\tablenotetext{a}{Relative to $t_0$ (MJD 60578.3).}
 \label{tab:thatcher_phot_all}
\end{deluxetable*}

\begin{deluxetable*}{ccccccc}
\tablecaption{Optical Spectroscopy \label{tab:optical_wpp_spec}}
\tablewidth{0pt}
\tablehead{
\colhead{UTC Date} & 
\colhead{MJD} &
\colhead{Phase\tablenotemark{a}} &
\colhead{Telescope} & 
\colhead{Instrument} & 
\colhead{Grating (Blue \& Red) } &
\colhead{Exposure Time (s; Blue / Red)} 
}
\decimalcolnumbers
\startdata
2024-09-29 & 60582.4 & 4.1 & Shane &  Kast & 600 / 4310 \& 600 / 7500 & 2460 / 2400
\\
2024-10-01 & 60584.5 & 6.2 & Shane & Kast & 452 / 3306 \& 600 / 7500 & 2460 / 2400
\\
2024-10-01 & 60584.9 & 6.6 & SALT & RSS &  PG0900 / 5100 \& PG0900 / 5700 & 1200 / 1200
\\
2024-10-03 & 60586.4 & 8.1 & Shane & Kast & 452 / 3306 \& 300 / 7500  & 1870 / 1800 
\\
2024-10-04 & 60586.5 & 9.1 & Shane & Kast &    600 / 4310 \& 300 / 7500  &  1835  / 1800 
\\
2024-10-05 & 60588.4 & 10.1 & Shane & Kast &  600 / 4310 \& 300 / 7500 & 1530 / 1500
\\
2024-10-05 & 60588.5 & 10.2 & Keck I & LRIS &  600 / 4000 \& 400 / 8500   &   600 /   600 
\\
2024-10-07 & 60590.4 & 12.1 & Keck I & LRIS & 400 / 3400 \& 400 / 8500 & 300 / 300
\\
2024-10-10 & 60593.3 & 15.0 & DeVeny & LDT & 300 / 5800  & 1200
\\
2024-10-11 & 60594.4 & 16.1 & Shane & Kast & 600 / 4310 \& 300 / 7500 & 10980 / 10800
\\
2024-10-12 & 60595.4 & 17.1 & Shane & Kast & 452 / 3306 \& 300 / 7500 &  3090 / 3000 
\\
2024-10-24 & 60607.4 & 29.0 & Shane & Kast & 452 / 3306 \& 300 / 7500 & 4290 / 4200
\\
2024-10-25 & 60608.9 & 30.6 & SALT & RSS & PG0900 / 5100 \& PG0900 / 5700 & 1200 / 1200
\\ 
2024-10-30 & 60613.4 & 35.1 & Keck I & LRIS & 600 / 4000 \& 400 / 8500  &  2400  / 2400 
\\
2024-11-08 & 60622.4 & 44.1 & Keck I & LRIS & 600 / 4000 \& 400 / 8500 & 4500 / 4500
\\
2024-11-27 & 60641.3 & 63.0 & Keck I & LRIS & 600 / 4000 \& 400 / 8500 & 5400 / 5250
\enddata
\tablenotetext{a}{Relative to $t_0$ (MJD 60578.3).}

\end{deluxetable*}

\begin{deluxetable*}{ccccccc}
\tablecaption{NIR Spectroscopy} 
\tablewidth{0pt}
\tablehead{
\colhead{UTC Date} & 
\colhead{MJD} &
\colhead{Phase\tablenotemark{a}} &
\colhead{Telescope} & 
\colhead{Instrument} & 
\colhead{Filter}  &
\colhead{Exposure Time (s; $JH$ / $HK$)}
}
\decimalcolnumbers
\startdata
2024-10-05 & 60588.3 & 10.0  & Gemini & F2 & 
$JH+HK$ & 840 / 840 \\
2024-10-16 & 60599.2 & 20.9 & Gemini & F2 &
$JH$ & 600 \\
2024-10-17 & 60600.2 & 21.9 & Gemini & F2 & 
$JH+HK$ & 960 / 2040\\
2024-10-19 & 60602.5 & 24.2 & Keck II & NIRES & & 2200 
\enddata
\tablenotetext{a}{Relative to $t_0$ (MJD 60578.3).}
\label{tab:nir_wpp_spec}
\end{deluxetable*}

\bibliography{AT2024wpp_v1}{}
\bibliographystyle{aasjournal}

\end{document}